\documentclass[onecolumn, a4paper, accepted=2023-10-19]{quantumarticle}
\pdfoutput=1
\usepackage[utf8]{inputenc}
\usepackage[numbers]{natbib}
\usepackage[bookmarks=false,colorlinks=true,urlcolor=blue,citecolor=blue,linkcolor=blue]{hyperref}
\usepackage{booktabs}
\usepackage{qcircuit}
\usepackage{amsmath,amssymb,mathtools}
\usepackage{graphicx}
\usepackage{bm,color}
\usepackage{xcolor}
\usepackage[capitalise]{cleveref}
\usepackage{multirow}
\usepackage{amsfonts} 
\usepackage{blkarray}
\usepackage{bbold}
\usepackage{tikz}
\usepackage{physics,braket}
\crefname{figure}{Fig.}{Figs.}

\usepackage{subcaption}

\newcommand{\be}{\begin{equation}}
\newcommand{\ee}{\end{equation}}

\newcommand{\lf}{\left}
\newcommand{\rt}{\right}

\newcommand{\sket}[1]{| #1 \rangle}

\newcommand{\czerory}[3]{\mathrm{C_0R}_{Y, #1, #2} (#3)}

\definecolor{nicepurple}{rgb}{0.55, 0.15, 0.75}
\definecolor{FNALGreen}{RGB}{76, 140, 43}
\definecolor{lightkblue}{RGB}{0, 0, 216}
\definecolor{deepsaffron}{rgb}{1.0, 0.6, 0.2}
\definecolor{sohaib}{rgb}{20, 20, 20}

\DeclarePairedDelimiter{\ceil}{\lceil}{\rceil}
\definecolor{brickred}{rgb}{0.80, 0.25, 0.33}

\newcommand{\mathedit}[1]{ \begingroup \color{blue} #1 \endgroup }

\begin{document}

\title{Preparing quantum many-body scar states on quantum computers}

\author{Erik J.~Gustafson}
\affiliation{Superconducting Quantum Materials and Systems Center (SQMS), Fermi National Accelerator Laboratory, Batavia, IL 60510, USA}
\affiliation{Fermi National Accelerator Laboratory, Batavia, IL, 60510, USA}

\author{Andy C.~Y.~Li}
\affiliation{Superconducting Quantum Materials and Systems Center (SQMS), Fermi National Accelerator Laboratory, Batavia, IL 60510, USA}
\affiliation{Fermi National Accelerator Laboratory, Batavia, IL, 60510, USA}

\author{Abid Khan}
\affiliation{Superconducting Quantum Materials and Systems Center (SQMS), Fermi National Accelerator Laboratory, Batavia, IL 60510, USA}
\affiliation{Department of Physics, University of Illinois Urbana-Champaign, Urbana, IL, United States 61801}
\affiliation{USRA Research Institute for Advanced Computer Science (RIACS), Mountain View, CA, 94043, USA}
\affiliation{Quantum Artificial Intelligence Laboratory (QuAIL), NASA Ames Research Center, Moffett Field, CA, 94035, USA}

\author{Joonho Kim}
\affiliation{Superconducting Quantum Materials and Systems Center (SQMS), Fermi National Accelerator Laboratory, Batavia, IL 60510, USA}
\affiliation{Rigetti Computing, Berkeley, CA, 94710, USA}

\author{Do\~ga Murat K\"urk\c{c}\"uo\~glu}
\affiliation{Superconducting Quantum Materials and Systems Center (SQMS), Fermi National Accelerator Laboratory, Batavia, IL 60510, USA}
\affiliation{Fermi National Accelerator Laboratory, Batavia, IL, 60510, USA}

\author{M. Sohaib Alam}
\affiliation{Superconducting Quantum Materials and Systems Center (SQMS), Fermi National Accelerator Laboratory, Batavia, IL 60510, USA}
\affiliation{USRA Research Institute for Advanced Computer Science (RIACS), Mountain View, CA, 94043, USA}
\affiliation{Quantum Artificial Intelligence Laboratory (QuAIL), NASA Ames Research Center, Moffett Field, CA, 94035, USA}

\author{Peter P.~Orth}
\affiliation{Superconducting Quantum Materials and Systems Center (SQMS), Fermi National Accelerator Laboratory, Batavia, IL 60510, USA}
\affiliation{Department of Physics and Astronomy, Iowa State University, Ames, IA 50011, USA}
\affiliation{Ames National Laboratory, Ames, IA 50011, USA}
\affiliation{Department of Physics, Saarland University, 66123 Saarbr\"ucken, Germany}

\author{Armin Rahmani}
\affiliation{Department of Physics and Astronomy and Advanced Materials Science and Engineering Center, Western Washington University, Bellingham, WA 98225, USA}

\author{Thomas Iadecola}
\email{iadecola@iastate.edu}
\affiliation{Superconducting Quantum Materials and Systems Center (SQMS), Fermi National Accelerator Laboratory, Batavia, IL 60510, USA}
\affiliation{Department of Physics and Astronomy, Iowa State University, Ames, IA 50011, USA}
\affiliation{Ames National Laboratory, Ames, IA 50011, USA}
%\date{\today}

\begin{abstract}
Quantum many-body scar states are highly excited eigenstates of many-body systems that exhibit atypical entanglement and correlation properties relative to typical eigenstates at the same energy density. Scar states also give rise to long-lived coherent dynamics when the system is prepared in a special initial state having finite overlap with them. Many models with exact scar states have been constructed, but the fate of scarred eigenstates and dynamics when these models are perturbed is difficult to study with classical computational techniques. In this work, we propose state preparation protocols that enable the use of quantum computers to study this question. We present protocols both for individual scar states in a particular model, as well as superpositions of them that give rise to coherent dynamics. For superpositions of scar states, we present both a system-size-linear depth unitary and a finite-depth nonunitary state preparation protocol, the latter of which uses measurement and postselection to reduce the circuit depth. For individual scarred eigenstates, we formulate an exact state preparation approach based on matrix product states that yields quasipolynomial-depth circuits, as well as a variational approach with a polynomial-depth ansatz circuit. We also provide proof of principle state-preparation demonstrations on superconducting quantum hardware.

\end{abstract}

    \maketitle

\section{Introduction}

Recent decades have seen remarkable advances in our understanding of how quantum statistical mechanics can emerge from isolated strongly-interacting quantum mechanical systems. 
One of the most fundamental of these advances is the so-called eigenstate thermalization hypothesis (ETH)~\cite{Deutsch91,Srednicki94,D'Alessio16,Deutsch18}, which posits that individual quantum mechanical eigenstates at finite energy density become locally equivalent in the thermodynamic limit to equilibrium Gibbs ensembles at a corresponding temperature. Such eigenstates also govern the approach to this local equilibrium under unitary dynamics~\cite{Rigol08}. In parallel with these developments, there have been enormous strides towards realizing quantum technologies based on coherent quantum systems that are approximately isolated on experimentally relevant time scales. This gives rise to the possibility of testing the ETH and the related phenomenon of quantum information scrambling experimentally using analog quantum simulators~\cite{Kaufman16,Gross17,Monroe21,Zhu22,Wang22} and digital quantum computers~\cite{Mi21}.

Along with this progress in understanding the ETH and quantum thermalization has come the realization that there are quantum systems that do not thermalize under certain conditions. Two notable examples include integrable~\cite{Polkovnikov11,Vidmar16} and many-body localized systems~\cite{Nandkishore15,Altman15,Abanin19}, where an extensive number of conserved quantities preclude the possibility of reaching a conventional locally thermalized state. An alternative means of avoiding thermalization is provided by quantum many-body scars (QMBS), a phenomenon whereby nonintegrable quantum systems exhibit a set of rare finite-energy-density eigenstates that do not satisfy the ETH~\cite{SerbynReview:2021,MoudgalyaReview:2021,Chandran22}. Such eigenstates have been found in a variety of contexts, including the Affleck-Kennedy-Lieb-Tasaki (AKLT) model~\cite{Moudgalya18a,Moudgalya18b}, ensembles of Rydberg atoms~\cite{Bernien17,Turner18a,Turner18b,Bluvstein21}, and various other interacting spin~\cite{Schecter19,Iadecola20,ODea20,Pakrouski20,Moudgalya20b,Ren21,Tang21,Ren22,Langlett22}, bosonic~\cite{Wildeboer22, su2023observation}, and fermionic~\cite{Mark20,Moudgalya20,Pakrouski21,PhysRevB.107.L201105,PhysRevB.107.205112, van2022dynamical, osborne2023spin} models. These eigenstates can also give rise to coherent periodic dynamics from certain initial states, which has allowed QMBS to be observed in quantum simulation experiments~\cite{Bernien17,Bluvstein21,su2023observation,Zhang22}.

Substantial effort has been devoted to formulating mathematical criteria for the emergence of QMBS~\cite{ODea20,Pakrouski20,Ren21,Langlett22,Wildeboer22,Moudgalya22}, and all examples so far require some degree of fine tuning. Understanding the fate of scarred eigenstates and the associated coherent dynamics under perturbations is thus an important research direction, but relatively little progress has been made so far. Ref.~\cite{Lin19} found a general lower bound on the thermalization timescale for a scarred state in the presence of a perturbation, $t_* = O(\epsilon^{-1/(d+1)})$, where $\epsilon$ is the perturbation strength and $d$ is the system's spatial dimension. However, this bound relies only on the underlying Hamiltonian's spatial locality, and numerical studies of specific models in, e.g., Refs.~\cite{Lin19,Langlett22} found lifetimes that substantially exceed this bound. 

Studying the lifetime of QMBS under perturbations is challenging because, although scarred eigenstates often have modest entanglement and can be represented efficiently in terms of matrix product states (MPSs)~\cite{Moudgalya18b,Moudgalya20b,Zhang22b}, perturbations couple them to states nearby in energy which typically have extensive volume-law entanglement. Analytical perturbation theory is impractical here owing both to the complexity of these highly excited eigenstates and to the exponentially large density of states at finite energy density. Numerical methods to directly evaluate the real time evolution of a scarred state under perturbations also encounter challenges. Exact methods are limited to small system sizes, while approximate tensor network methods~\cite{Schollwock11,Orus14} are generally limited to early times~\cite{Luitz17}.

A natural question is whether one could exploit quantum computers to investigate the behavior of scarred states under perturbations. It has long been known that quantum computers can evaluate real-time dynamics efficiently~\cite{Lloyd96}; indeed, simulating quantum dynamics is one of the leading candidates for near-term practical quantum advantage~\cite{Childs18,Daley22}. However, even if we assume access to a noiseless quantum computer that can perform accurate time evolution, we are still left with the challenge of state preparation: how can we prepare scarred states on a digital quantum computer, and what are the resources required to do so? This is the subject of the present work.

There are two general state preparation tasks that one faces in this context, depending on whether one wants to investigate the lifetime of scarred \textit{dynamics} or scarred \textit{eigenstates}. In the first case, the aim is to time-evolve a \textit{superposition} of scarred states that exhibits periodic dynamics in the unperturbed limit and extract the lifetime of the observed oscillations. In many cases of interest, a product state is sufficient for these purposes and the state preparation is therefore trivial~\cite{Bernien17,Turner18a,Schecter19,Bluvstein21,su2023observation,Zhang22,Langlett22,Wildeboer22,Chen22}. However, in other cases, the simplest superposition of scar states is area-law entangled and has a nontrivial MPS representation with a finite correlation length~\cite{Chattopadhyay19,Iadecola20,Mark20b}, and here some thought must be put into the most efficient method to prepare such states.

In the second case, the aim is to prepare an \textit{individual} scarred eigenstate and evolve it under the perturbed Hamiltonian. In many cases of interest, scarred eigenstates have entanglement scaling logarithmically with system size~\cite{Vafek17,Moudgalya18b,Turner18b,Schecter19,Iadecola20,Chattopadhyay19,Mark20,Moudgalya20}, similar to critical states in one dimension that are described by conformal field theories. The entanglement content of these states is modest compared to typical volume-law states at the same energy density, suggesting the possibility of an efficient state preparation circuit. In fact, in many of the simplest examples of QMBS the scarred eigenstates are equivalent to Dicke states~\cite{Choi18,Schecter19,Mark20,Moudgalya20,Langlett22,Wildeboer22}, for which a polynomial-depth state preparation circuit is known~\cite{Bartschi19}. However, in other examples, the states of interest can be viewed as projected or finite-momentum Dicke states~\cite{Iadecola20,Chattopadhyay19,Tang21}, about which much less is known.
Thus in general it is an open question what are the minimal quantum resources needed to prepare scarred eigenstates.

In this work we address both of the above cases for a specific model with QMBS. In Sec.~\ref{sec:model} we define the model and state preparation tasks in more detail. In Sec.~\ref{sec: xiprep} we consider the problem of preparing a particular class of superpositions of scarred eigenstates that can be realized as a one-parameter family of MPSs with bond dimension $\chi=2$. We consider two related approaches. First, we explicitly construct a linear-depth circuit that prepares the desired state with perfect fidelity. Second, we discuss a probabilistic method that prepares the desired states in constant depth using measurements and postselection. The latter method has a postselection success probability that decays exponentially with system size, albeit with a base that can be tuned by adjusting the circuit depth. This allows for a flexible tradeoff between circuit depth and success probability that is advantageous for implementation on near-term quantum hardware. In Sec.~\ref{sec: Skprep}, we discuss two strategies for the preparation of individual scarred eigenstates. In the first strategy, we identify MPS representations for the scarred eigenstates and convert these to quantum circuits that prepare the states with perfect fidelity in quasipolynomial depth. In the second, we propose a polynomial-depth variational ansatz that we show numerically captures the scarred eigenstates with at least $99\%$ fidelity at numerically accessible system sizes. In both Secs.~\ref{sec: xiprep} and \ref{sec: Skprep}, we provide proof-of-concept demonstrations of the respective state preparation tasks on Rigetti and IBM quantum processing units (QPUs), often with some additional simplifications in order to obtain better results on hardware. Finally, we provide a conclusion and outlook in Sec.~\ref{sec: conclusion}. In the appendices, we discuss an alternative state preparation protocol based on adiabatic evolution (Appendix~\ref{sec: Adiabatic}), details about state preparation circuit derivations (Appendix~\ref{sec:stochderivation}), and additional QPU results on IBM hardware (Appendix~\ref{sec:ibm_result}).

\section{Model and State Preparation Tasks}
\label{sec:model}

In this work, we select a particular reference model with QMBS to exemplify our state preparation techniques for scarred eigenstates and their superpositions. We consider the spin-1/2 model defined in Ref.~\cite{Iadecola20}, whose Hamiltonian reads
\begin{align}
\label{eq:H0}
\begin{split}
    H_0&=\lambda\sum^{N-1}_{i=2}(X_i-Z_{i-1}X_iZ_{i+1})+\Delta \sum^N_{i=1} Z_i + J \sum^{N-1}_{i=1}Z_{i}Z_{i+1}.
\end{split}
\end{align}
This Hamiltonian acts on a chain of $N$ qubits with open boundary conditions, and each qubit is equipped with Pauli operators $X_i,Y_i,$ and $Z_i$. Note that $H_0$ commutes with the operators $Z_1$ and $Z_N$---the $Z$-basis projections of the edge qubits are therefore conserved quantities. $H_0$ also conserves the number of Ising domain walls, measured by the operator $n_{\rm DW}=\sum^{N-1}_{i=1}(1-Z_iZ_{i+1})/2$. The $\lambda$ term in Eq.~\eqref{eq:H0} can be viewed as a kinetic term for domain walls, while the Ising interaction $J$ serves as a chemical potential for the domain walls. The $\Delta$ term induces nonlocal interactions between domain walls and makes the model nonintegrable. It is interesting to note that this model is dual to a $\mathbb Z_2$ lattice gauge theory coupled to fermionic matter~\cite{Borla20} and can be realized in Rydberg atom quantum simulators in the antiblockade regime~\cite{Ostmann19}.

The model \eqref{eq:H0} has two towers of QMBS states related by the global spin-flip operation $G=\prod^N_{i=1}X_i$. The first tower is given by
\begin{align}
\label{eq:Sk}
    \ket{\mathcal S_k}= \frac{1}{k!\sqrt{\mathcal N(N,k)}}(Q^\dagger)^k\ket{\Omega},
\end{align}
where $\mathcal N(N,k)=\binom{N-k-1}{k}$, $\ket{\Omega}=\ket{0\dots0}$, and $k=0,\dots,N/2-1$ (we take $N$ even for simplicity). The raising operator for this tower of states is given by
\begin{align}
\label{eq:Qdag}
    Q^\dagger = \sum^{N-1}_{i=2}(-1)^i P_{i-1}\sigma^+_i P_{i+1},
\end{align}
where $P_j = \ket 0_j\bra 0_j=(1+Z_j)/2$ and $\sigma^\pm_j = (X_j\mp i Y_j)/2$. The states $\ket{\mathcal S_k}$ are eigenstates of $H_0$ with energies $E_k = \Delta N + J(N-1)-(2\Delta+4J)k$. The second tower of states \begin{align}
\ket{\mathcal S^\prime_k}=G\ket{\mathcal S_k},\indent k=0,\dots,\frac{N}{2}-1    
\end{align}
has energies $E^\prime_k = -\Delta N + J(N-1)+(2\Delta-4J)k$. Both towers of states therefore have an extensive energy bandwidth, so that typical states in each tower correspond to highly excited states of the model $H_0$. Nevertheless, states in either tower for which $k/N$ is finite as $N\to\infty$ have bipartite entanglement entropy scaling as $\ln(N)$, in contrast to the volume-law entanglement entropy of typical eigenstates at the same energy density~\cite{Iadecola20}. In the remainder of the paper we will restrict our attention to the tower $\{\ket{\mathcal S_k}\}$, but all results we obtain for this tower hold equally well for the tower $\{\ket{\mathcal S^\prime_k}\}$.

Dynamical consequences of these scar states can be observed by evolving the system from a suitable family of initial states with support on the towers of interest. In Ref.~\cite{Iadecola20}, it was shown that the following family of states parameterized by $\xi\in\mathbb C$ is exclusively supported on the tower $\{\ket{\mathcal S_k}\}$:
\begin{align}
\label{eq: xi}
\ket{\xi} = \frac{1}{\sqrt{Z(|\xi|^2)}}\mathcal P_{\rm fib} \prod^{N-1}_{i=2}[1+(-1)^i\xi\sigma^+_i]\ket{\Omega},
\end{align}
where the projection operator
\begin{align}
\label{eq:pfib}
\mathcal P_{\rm fib}=\prod^{N-1}_{i=1}(1-P^\prime_{i}P^\prime_{i+1}),\indent P^\prime_i = 1-P_i
\end{align}
excludes any computational basis state in Eq.~\eqref{eq: xi} containing the local configuration $\ket{1}_i\ket{1}_{i+1}$, and where the normalization factor
\begin{align}
\label{eq:Z}
    Z(|\xi|^2)=\sum^{N/2-1}_{k=0}|\xi|^{2k}\mathcal N(N,k).
\end{align}
The state $\ket{\xi}$ can be decomposed onto the tower $\{\ket{S_k}\}$ as follows:
\begin{align}
\label{eq:xi_Sk}
    \ket{\xi}=\sum^{N/2-1}_{k=0}\xi^k\sqrt{\frac{\mathcal N(N,k)}{Z(|\xi|^2)}}\ket{\mathcal S_k}.
\end{align}
As such, for any value of $\xi$, evolving the initial state $\ket{\xi}$ under $H_0$ yields perfect periodic revivals of $\ket{\xi}$ with period $T=\pi/(\Delta+2J)$ set by the energy spacing between consecutive states in the tower. Unlike typical states in the tower $\{\ket{\mathcal S_k}\}$, the state $\ket{\xi}$ is area-law entangled. In fact, it can be written as an MPS with bond dimension $\chi=2$~\cite{Iadecola20}, so its bipartite entanglement entropy is upper bounded by $\ln 2$.

The state $\ket{\xi}$ has connections to several interesting problems in condensed matter and atomic physics. For instance, it is unitarily equivalent to a state found in Ref.~\cite{Lesanovsky12a} to be a good approximation to the ground state of a system of Rydberg atoms in the so-called Rydberg blockade regime~\cite{Jaksch00,Lukin01}, where the computational basis states $\ket{0}_i$ and $\ket{1}_i$ correspond to atom $i$ being in its ground state or a highly excited Rydberg state, respectively. Due to strong interactions, such systems are subjected to an energetic penalty for having two excited atoms next to one another. This constraint, sometimes called the ``Fibonacci constraint" because the number of states of $m$ qubits that satisfy it is given by $F_{m+2}$ (where $F_\ell$ is the $\ell$-th Fibonacci number), is implemented by the projector $\mathcal P_{\rm fib}$ in Eq.~\eqref{eq: xi}. Intriguingly, this constraint also emerges in theoretical descriptions of the $\nu=1/3$ Laughlin fractional quantum Hall (FQH) state in a particular quasi-1D limit~\cite{Nakamura12,Moudgalya19,Rahmani20,Kirmani22}. In fact, the ground state of the system in this limit is unitarily equivalent to $\ket{\xi}$ for a particular choice of the parameter $\xi$~\cite{Nakamura12}. Thus, our state preparation results for $\ket{\xi}$ will also be applicable to the seemingly disparate settings of Rydberg-atom quantum simulators and FQH liquids.

Studying the stability of scarred eigenstates and dynamics on quantum computers requires algorithms for high-fidelity preparation of the states $\ket{\mathcal S_k}$ and $\ket{\xi}$, respectively. This paper provides a survey of approaches to both state preparation tasks and a snapshot of their feasibility with current quantum hardware. We first address the preparation of the superposition state $\ket{\xi}$ in Sec.~\ref{sec: xiprep}, before moving onto the preparation of the scar states $\ket{\mathcal S_k}$ in Sec.~\ref{sec: Skprep}.

The alternating signs $(-1)^i$ appearing in Eqs.~\eqref{eq:Qdag} and \eqref{eq: xi} can be removed by the simple unitary circuit $\prod_{i\text{ odd}}Z_i$. Therefore, it is useful to define the states
\begin{subequations}
\label{eq:tilde}
\begin{align}
    |\tilde\xi\rangle = \left(\prod_{i\text{ odd}}Z_i\right)\ket{\xi}
\end{align}                    
and
\begin{align}
    |\tilde{\mathcal S}_k\rangle = \left(\prod_{i\text{ odd}}Z_i\right)\ket{\mathcal S_k},
\end{align}
\end{subequations}
which are now equal-amplitude and equal-sign superpositions of computational basis states. We will at times find it more convenient to work with these ``tilde" states than with the original states.

Before proceeding, we highlight that preparing scarred eigenstates and superpositions in our reference model~\eqref{eq:H0} is particularly challenging due to the Fibonacci constraint. 
In many scarred models of interest, the scarred eigenstates can be realized as Dicke states~\cite{Choi18,Schecter19,Mark20,Moudgalya20,Langlett22,Wildeboer22}, for which a polynomial-depth circuit construction is known~\cite{Bartschi19}. 
Likewise, low-entanglement superpositions of scar states in these models can be realized as product states, for which state preparation is trivial. Our motivation in selecting this more complicated model is that kinetic constraints and the more elaborate entanglement structures they enable are present in a variety of other models, including the AKLT model~\cite{Moudgalya18a,Moudgalya18b,Mark20b,Chattopadhyay19}. Thus, grappling with such constraints and assessing their impact on state preparation complexity is necessary in order to study the full variety of QMBS on quantum computers.

\section{Preparing the State $\ket{\xi}$}
\label{sec: xiprep}

There are several possible approaches to preparing the superposition state $\ket{\xi}$ from  Eq.~\eqref{eq: xi}. In Ref.~\cite{Iadecola20} it was shown that $\ket{\xi}$ is the unique ground state of a local parent Hamiltonian with finite correlation length (see also Ref.~\cite{Lesanovsky12a}). Thus one approach is to prepare $\ket{\xi}$ adiabatically using an appropriate parameter sweep from a zero-correlation-length paramagnetic Hamiltonian. We investigate this approach in App.~\ref{sec: Adiabatic}, where we find evidence that the gap of the interpolating Hamiltonian does not close with increasing $N$ and the state $\ket{\xi}$ is a suitable candidate for adiabatic state preparation. In practice, adiabatic state preparation on a digital quantum computer suffers from Trotter error, even for a fully gapped interpolation Hamiltonian and assuming perfect implementation of the Trotter circuit. For this reason a finite depth circuit always incurs finite error. This motivates the consideration of alternative state preparation strategies. 

In Sec.~\ref{sec:Linear}, we demonstrate that perfect state preparation can be achieved with a circuit of depth $O(N)$. Sec.~\ref{subsecn:prob-prep} shows that a stochastic strategy \cite{Hubisz:2020vhx,foss-feig_holographic_2020,tantivasadakarn_long-range_2022,lu_measurement_2022,Friedman22,Smith22} using measurements and post-selection can reduce the circuit depth to a constant at the price of an exponential post-selection overhead. Finally, in Sec.~\ref{sec:xi-qpu}, we benchmark both state preparation strategies on Rigetti QPUs. We show additional QPU results from IBM hardware in Appendix~\ref{sec:ibm_result}.

\subsection{Linear-Depth Unitary Circuit}
\label{sec:Linear}

To facilitate our discussion, we rewrite the superposition state $\ket{\xi}$ with open-boundary conditions on $N$ qubits [Eq.~\eqref{eq: xi}] as
\begin{align}
	\label{eq:xi_m_state_definition}
	\ket{\xi} = \ket 0\otimes \ket{\xi;N-2} \otimes \ket0.
\end{align}
In this section, we show that the state $\ket{\xi;m}$ as well as its counterpart $|\tilde\xi;m\rangle$ with alternating phases removed per Eqs.~\eqref{eq:tilde}
can be prepared in linear depth $O(m)$ by a unitary circuit $\mathfrak{U}_{\xi}(m)$ consisting of $(m-1)$ controlled $Y$-rotation gates and one $Y$-rotation gate. We note that a similar linear depth circuit was obtained in the FQH context in Ref.~\cite{Rahmani20}, but that the nonunitary approach explored in Sec.~\ref{subsecn:prob-prep} has not been discussed in this context.

\begin{figure}
\centering
\includegraphics[width=0.5\linewidth]{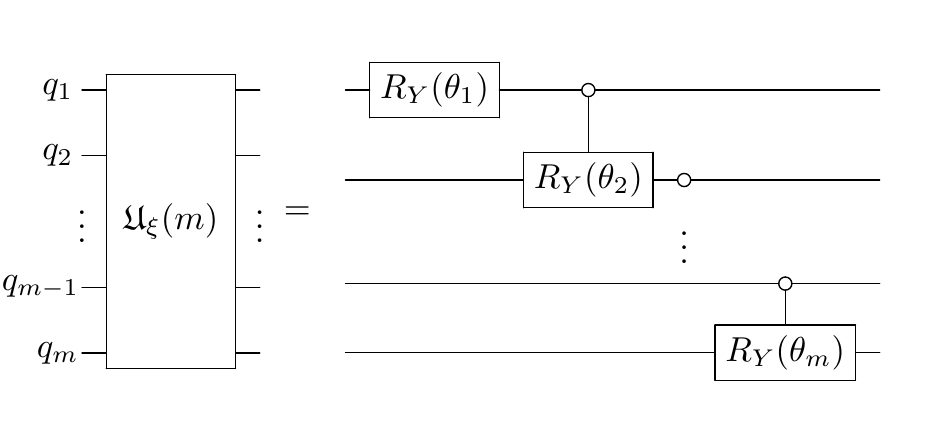}
\caption{Linear-depth circuit showing the preparation gate $\mathfrak{U}_{\xi}(m)$ to prepare $\ket{\xi;m}$ from the zero state $\ket{\Omega}$.  circuit consists of $(m-1)$ control-rotation gates and one rotation gate. The rotation angles $\theta_j$ are given in \cref{eq:linear_preparation_circuit_recursive}.}
\label{fig:linear_depth_preparation_circuit}
\end{figure}

To understand this preparation circuit, we recall that $\ket{\xi}$ or $\ket{\xi;m}$ is the superposition of all computational basis states excluding any local configuration $\ket{1}_{j-1} \ket{1}_{j}$ with the weights of each state determined by the parameter $\xi$.
Starting with all qubits in the $\ket 0$ state, the absence of $\ket{1}_{j-1} \ket{1}_{j}$ pairs is guaranteed under a sequence of controlled $Y$-rotations $\czerory{j-1}{j}{\theta_j}$ by an angle $\theta_j$ targeted on the $j$-th qubit that is triggered only by $\ket{0}$ of the $(j-1)$-th qubit.
For a set of properly chosen rotation angles $\theta_j$, $\ket{\xi;m}$ can be prepared by this preparation circuit:
\begin{align}
\begin{split}
\ket{\xi;m} = & \ \mathfrak{U}_{\xi}(m)\ket{\Omega}, \\
\mathfrak{U}_{\xi}(m) = & \lf[\prod_{j=2}^{m} \czerory{j-1}{j}{\theta_j} \rt] \mathrm{R}_{Y, 1}(\theta_1).
\end{split}
\label{eq:linear_preparation_circuit}
\end{align}
The corresponding circuit diagram of the preparation gate $\mathfrak{U}_{\xi}(m)$ is shown in \cref{fig:linear_depth_preparation_circuit}.
The angles are determined as a function of $\xi$ according to the following recursion relation:
\begin{align}
\begin{split}
\theta_j = & 2 \, \mathrm{arg}  \lf[ 1 + i \frac{(-1)^{j+1} \xi}{\phi_{j+1}}  \rt],~\phi_j = \sqrt{1 + \lf| \frac{(-1)^{j+1} \xi}{\phi_{j+1}} \rt|^2} \text{\ \ and \ \ } \phi_{m+1} = 1.
\end{split}
\label{eq:linear_preparation_circuit_recursive}
\end{align}
The preparation circuit $\mathfrak U_{\tilde\xi}(m)$ for the state $|\tilde\xi; m\rangle$ is obtained by simply removing the alternating phase factors $(-1)^{j+1}$ from the above definition.
%\remove{We will prove this recursion relation for the rotation angles and the preparation circuit in the following.}
%\edit{
The proof of the preparation circuit and the above recursion relation for the rotation angles is provided in \cref{sec:stochderivation}.
%}

\subsection{Probabilistic Constant-Depth Circuit with Postselection}
\label{subsecn:prob-prep}

While the linear-depth circuit explored in the previous section is sufficient for state preparation, it may be challenging to implement on near-term devices for large $N$ owing to the linear circuit depth.
Here, we show that it is possible to prepare the equal-amplitude superposition state $|\tilde \xi; N-2\rangle$ stochastically in constant depth using measurements and postselection. The idea is to  prepare $k$ $m$-site blocks in the state $|\tilde \xi;m\rangle$, which can be achieved in depth $O(m)$ using the circuit $\mathfrak{U}_{\tilde\xi}(m)$ [\cref{eq:linear_preparation_circuit} with alternating phases removed in the recursive formula \cref{eq:linear_preparation_circuit_recursive}]. The resulting state $|\tilde\xi;m\rangle^{\otimes k}$ obeys the Fibonacci constraint enforced by the projector $\mathcal P_{\rm fib}$ [\cref{eq:pfib}] within each $m$-site block, but adjacent blocks need not obey the constraint. To ``stitch" the adjacent blocks together into a state that globally satisfies the constraint, adjacent $m$-site blocks are each coupled to an ancilla qubit using an appropriate unitary operation. Measuring the ancilla qubit and postselecting onto an appropriate measurement outcome prepares the desired state $|\tilde\xi;km\rangle$, which can then be trivially converted to $\ket{\xi;km}$ using Eq.~\eqref{eq:tilde}.

To illustrate, let us set $\xi=1$ for simplicity (this is not strictly necessary but does simplify the analytical expressions for the states and success probabilities). As an example, consider stitching together two copies of the two-qubit state $\ket{1;2}$. The two-qubit state is given by
\begin{align}
|\tilde1;2\rangle = \frac{1}{\sqrt{3}}\Big(\ket{00} + \ket{10} + \ket{01}\Big),
\end{align}
and the state we aim to prepare is
\begin{align}
\begin{split}
    |\tilde 1;4\rangle&=\frac{1}{\sqrt{8}}(\ket{0000}+\ket{0001}+\ket{0010}+\ket{0100}+\ket{1000}+\ket{1010}+\ket{0101}+\ket{1001}).
\end{split}
\end{align}
We first initialize the system in $\ket{\Omega}=\ket{0000}$ and apply the circuit $\mathfrak{U}_1(2)$ to the two consecutive two-qubit blocks to prepare the state
\begin{align}
\begin{split}
|\tilde 1;2\rangle\otimes|\tilde 1;2\rangle = & \frac{1}{3}\Big(\ket{00} + \ket{10} + \ket{01}\Big)\otimes\Big(\ket{00} + \ket{10} + \ket{01}\Big).
\end{split}
\end{align}
There is exactly one configuration in the above superposition that would be projected away by $\mathcal P_{\rm fib}$, namely $\ket{0110}$. Therefore if we apply  a Toffoli ($\mathrm{CCNOT}$) gate controlled on qubits 2 and 3 and targeted on an ancilla qubit initialized in the state $\ket{0}_a$ to the above state, we obtain
\begin{align}
\frac{\sqrt 8}{3}|\tilde 1;4\rangle\ket{0}_a + \frac{1}{3}\ket{0110}\ket{1}_a.
\end{align}
Measuring the ancilla qubit in the computational basis, we obtain the desired state $|\tilde 1;4\rangle$ whenever the measurement outcome is $0$, which occurs with probability $8/9$.

The same procedure can be generalized to stitch together $k$ copies of $|\tilde 1;2\rangle$ into the state $|\tilde 1;2k\rangle$ using $k-1$ ancilla qubits initialized in the $\ket 0$ state. Each ancilla qubit is coupled to the array of $2k$ primary qubits using a Toffoli gate controlled by the neighboring qubits from consecutive two-site blocks. In this way, the state of each ancilla qubit records whether the Fibonacci constraint is violated for the pair of primary qubits to which it is coupled. The probability that the Fibonacci constraint is satisfied (such that the ancilla register is in the state $\ket{0\dots 0}_a$) is given in terms of Fibonacci numbers $F_\ell$ by $F_{2k+2}/F_4^k$, which decays exponentially with $k$.

\begin{figure}
\centering
\includegraphics[width=0.5\linewidth]{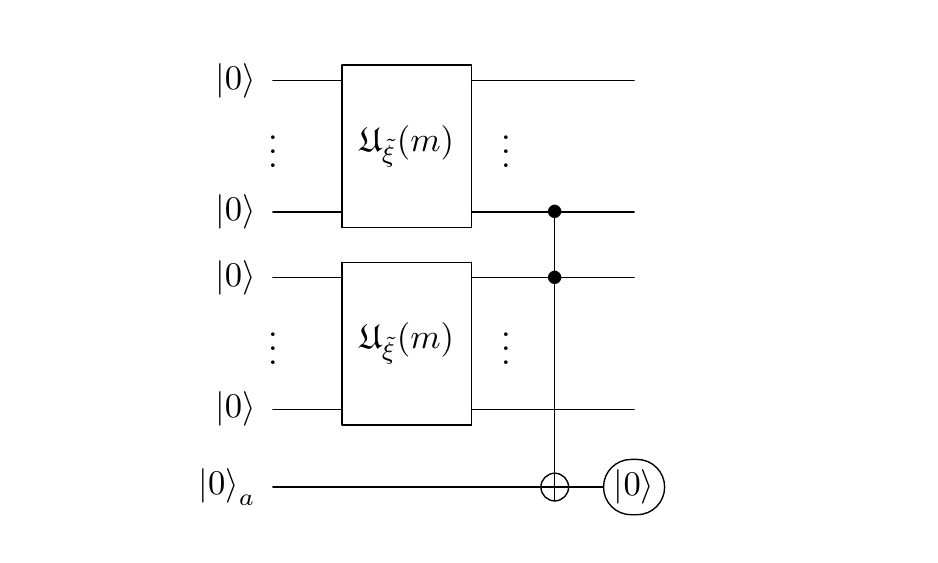}
\caption{Example circuit for stitching two copies of $|\tilde\xi;m\rangle$ together into $|\tilde\xi;2m\rangle$ using a single Toffoli gate targeted onto an ancilla, which is then measured and postselected onto the $\ket{0}$ state. The circuit $\mathfrak U_{\tilde\xi}(m)$ is simply Eq.~\eqref{eq:linear_preparation_circuit} with the alternating signs removed from the angles in Eq.~\eqref{eq:linear_preparation_circuit_recursive}.}
\label{fig:quantumcircuitstitchingxi}
\end{figure}

The above success probability can be improved by using the same strategy to stitch together larger blocks. For example, suppose we wish to prepare the state $|\tilde1;2m\rangle$ by stitching together two $m$-site blocks prepared in the state $|\tilde1;m\rangle$ using the $O(m)$-depth circuit $\mathfrak{U}_{\tilde1}(m)$. The stitching can again be achieved by coupling the two neighboring qubits from the two blocks to an ancilla qubit using a Toffoli gate.
An illustration of this procedure for general $\xi$ is shown in Fig.~\ref{fig:quantumcircuitstitchingxi}.
The success probability can be obtained by noting that there are $F_{2m+2}$ states of the full $2m$-qubit system that satisfy the Fibonacci constraint (for which a measurement of the ancilla qubit would yield $0$), while the initial tensor product state contains $F^2_{m+2}$ configurations; the success probability is then $F_{2m+2}/F_{m+2}^2$.

\begin{figure}[t]
\centering
\includegraphics[width=0.5\linewidth]{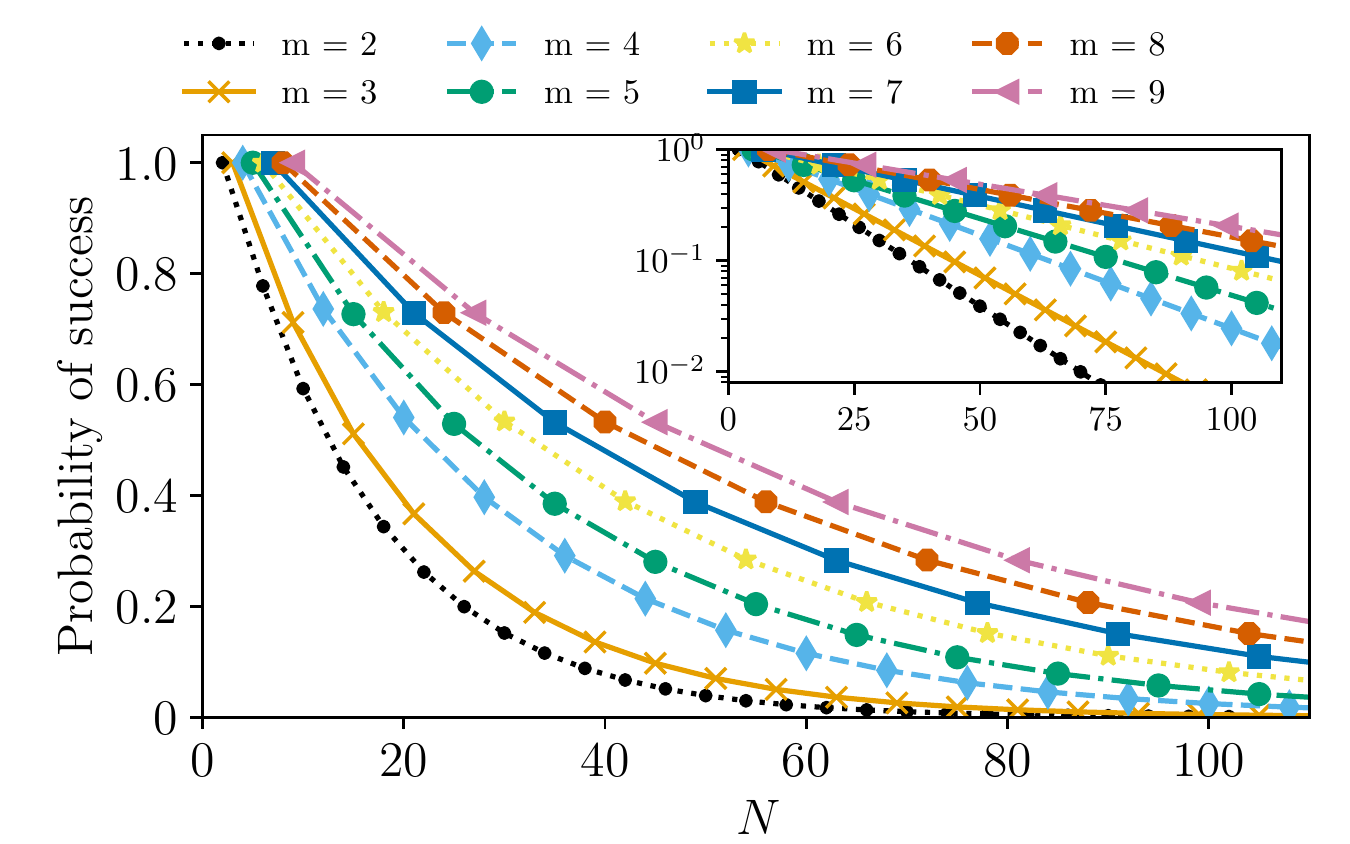}%{fig/test.pdf}
\caption{Success probability of the finite-depth postselection-based protocol for $N=km$ qubits and initial block size $m$. The inset shows the same data with a logarithmic scale on the $y$-axis, indicating that the success probability decays exponentially with $N$ with a base set by $m$; increasing $m$ increases the success probability by bringing the base closer to $1$.}
\label{fig:successprobabilitytheory}
\end{figure}

Applying the same logic to $k$ blocks of $m$ sites yields a postselection success probability
\begin{align}
\label{eq:success}
p_{\rm success}(m,k) = F_{km+2}/F_{m+2}^k\,,
\end{align}
which is plotted for various $m$ against $N=km$ in Fig.~\ref{fig:successprobabilitytheory}. Although this expression still decays exponentially with $k$ for fixed $m$, it grows with $m$ at fixed $k$. Thus, the exponential sampling overhead can be mitigated by increasing $m$ at the price of increasing the depth of the state preparation circuit.
Fig.~\ref{fig:successprobabilitytheory} shows the success probability for a lattice of length $N$ using blocks of length $m$. On present-day NISQ devices, gate error and qubit decoherence rates are sufficiently high that the reduction of circuit depth at the expense of postselection may be an acceptable tradeoff. We analyze this tradeoff further in Sec.~\ref{sec:xi-qpu} when we implement this state preparation protocol on quantum hardware.


It is natural to ask whether a deterministic (i.e., postselection-free) finite-depth state preparation protocol can be formulated in which unwanted ancilla measurement outcomes are corrected by local unitary operations instead of being discarded. Such a scheme can leverage the fact that the local form of the state in the vicinity of an ancilla measurement outcome of $1$ is fixed. For example, consider stitching together two states of the form $|\tilde 1; m\rangle$ with a single ancilla measurement, and suppose the measurement outcome was $1$. Then the primary qubit register is in the state
\begin{align}
    |\tilde1;m-2\rangle \otimes \ket{0110}\otimes|\tilde1;m-2\rangle,
\end{align}
since the state away from the central two sites (which are projected onto $\ket{11}$ by the measurement) obeys the Fibonacci constraint. One can imagine trying to correct the four site block $\ket{0110}$ using a unitary circuit controlled by the states of the two adjacent qubits from the $(m-2)$-site blocks to its left and right. However, the target states in the four cases (labeled by the four possible states of the adjacent qubits) have different normalization factors due to the Fibonacci constraint; therefore, the state of the qubit chain after feedback will not be an equal-amplitude superposition state. This situation should be contrasted with a recently proposed finite-depth deterministic scheme to prepare the AKLT ground state~\cite{Smith22}, which is able to correct undesired measurement outcomes with a local circuit. However, the correction operation designed in that work makes use of the fact that the AKLT state is a symmetry-protected topological state~\cite{Pollmann10,Pollmann12}, which is a property that is not shared by the state $\ket{\xi}$. We leave the question of whether an undesired measurement outcome in our probabilistic protocol can be corrected by a finite-depth circuit for future work.

\subsection{QPU Results}
\label{sec:xi-qpu}

\begin{figure*}
\centering
\centering
\includegraphics[width=0.485\textwidth]{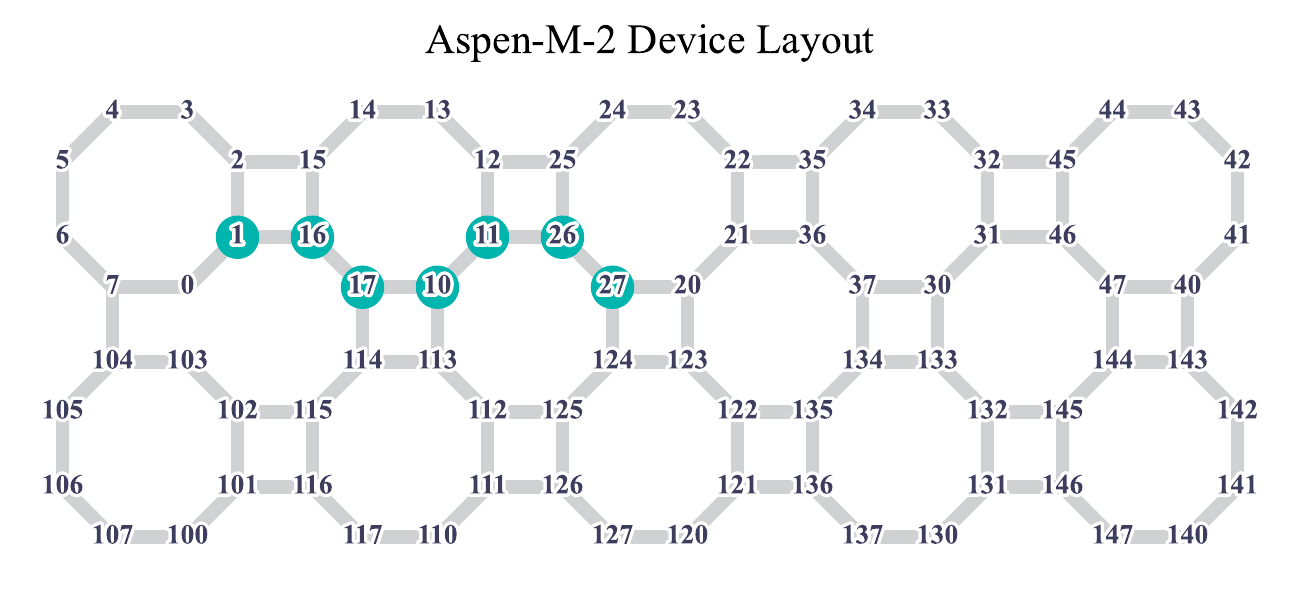}
\hfill
\includegraphics[width=0.485\textwidth]{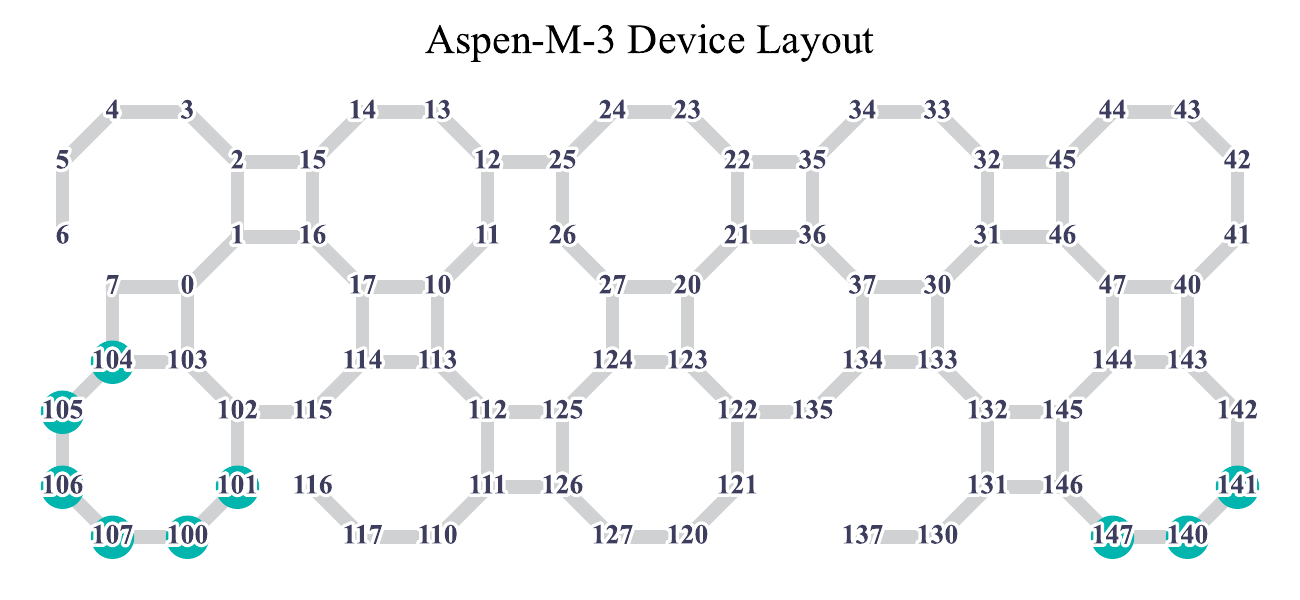}
\caption{Aspen M-2 and M-3 device layout graphs, with each node representing a superconducting qubit and each edge indicating connectivity via two-qubit gates. Circled nodes denote qubits used in the QPU experiments in Sec.~\ref{sec:xi-qpu}~and~\ref{sec:Sk_QPU}.}
\label{fig:aspen-layout}
\end{figure*}

We now implement the $|\xi\rangle$ state preparation protocols on Rigetti's Aspen QPUs (shown in Fig. \ref{fig:aspen-layout}) and on IBM hardware. We discuss the execution results of the linear-depth circuit in Sec.~\ref{sec:Linear} and its probabilistic post-selection variant in Sec.~\ref{subsecn:prob-prep}.
We set $\xi=1$ for numerical evaluations, in which case the Pauli rotation angles in Fig. \ref{fig:linear_depth_preparation_circuit}
can be summarized as
\begin{align}
\theta_{m-i+1} = 2 \tan^{-1}(\sqrt{F_{i+1} / F_{i}})
\end{align}
with Fibonacci coefficients $F_i = (1, 1, 2, 3, 5, 8, \cdots)_i$. 
The \texttt{quilc} compiler generates an optimized program for the instruction set architecture of QPU chips, compiling all logical gates into the following group of native gates: 
$$
\mathrm{RZ}(\theta), \mathrm{RX}(\pi/2), \mathrm{RX}(\pi), \mathrm{CPHASE}(\theta), \mathrm{CZ}, \mathrm{XY}(\theta).
$$

\begin{figure*}
\centering
\includegraphics[width=\textwidth]{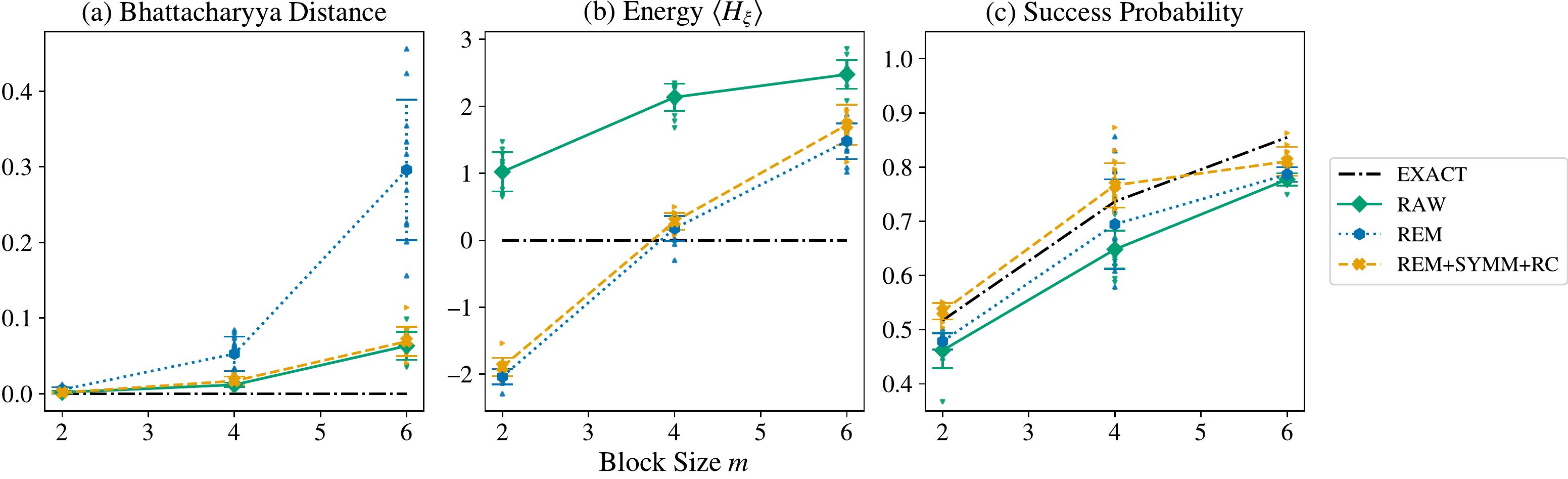}
\caption{
The QPU experiment results of the $|\xi=1\rangle$ state preparation circuit on Aspen-M devices, which uses the probabilistic post-selection protocol that combines multiple $m$-qubit blocks $\mathfrak{U}_{\xi}(m)$ to build the $N=14$ state. The indices of the participating Aspen qubits are: $\{105, 104\}$ (M-3) for $m=2$, $\{106, 107, 100, 101\}$ (M-3) for $m=4$, $\{16, 17, 10, 11, 26, 27\}$ (M-2) for $m=6$. 
Four distinct sets of measured values are presented: EXACT, RAW, REM, and REM+SYMM+RC, where REM, SYMM, and RC are acronyms for readout error mitigation, readout symmetrization \cite{Smith2021_readout_symm}, and randomized compilation \cite{Wallman2016_randomized_compiling}, respectively. 
The RAW samples are derived directly from the post-selection protocol using unprocessed bitstrings. On the other hand, the REM samples utilize corrected bitstrings, which undergo iterative Bayesian unfolding \cite{Nachman2020_readout_unfolding} prior to the post-selection. The REM+SYMM+RC samples are averages of REM outcomes over $30$ logically equivalent circuits compiled with Pauli twirling and readout symmetrization. For comparison, the noiseless expected values are also presented in the EXACT curve. Results in the middle and right panels are obtained with $10^4$ shots per sample and in the left panel with $10^5$ shots per sample. Specifically, an individual RC circuit contributes approximately $\frac{10^4}{30}$ or $\frac{10^5}{30}$ shots to the collected REM+SYMM+RC samples.
We gather $10$ sample data points for each quantity, varying the value of $m$. Error bars denote error of the mean over the different samples.
(a) Bhattacharyya distance, which is a measure of the difference between the ideal and measured bitstring probabilities versus block size $m$. (b) Energy versus $m$, which ideally vanishes. (c) Observed postselection success probability versus $m$. 
}

\label{fig:xi-state-n14}
\end{figure*}

\begin{figure*}
\centering
\includegraphics[width=\textwidth]{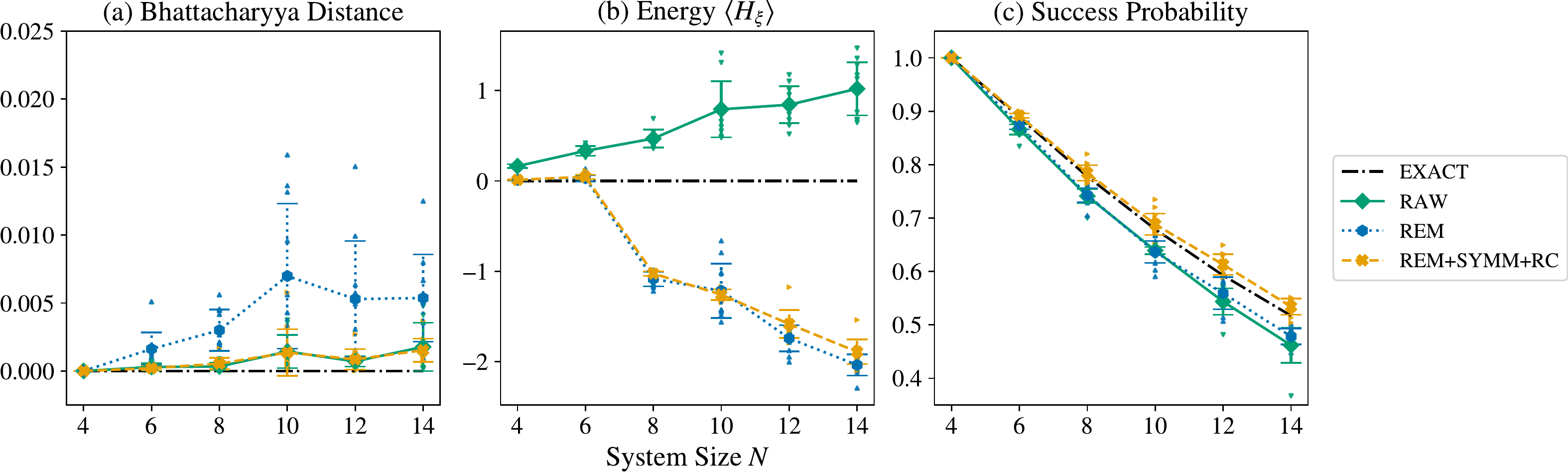}
\caption{The $|\xi=1\rangle$ state preparation experiment on Aspen M-3 that combines $\mathfrak{U}_{\xi}(m=2)$ to construct the $4 \leq N \leq 14$ states. The sample data points are obtained using the $\{105, 104\}$ device qubits. We refer to the caption of Figure~\ref{fig:xi-state-n14} for a detailed explanation of the data labels (EXACT, RAW, REM, REM+SYMM+RC) and the number of repeated measurements. We collect 10 sample points for each quantity, with the value of $N$ varying, where the error bars denote the error of the mean across different samples. (a) Bhattacharyya distance, a measure of the difference between the ideal and measured bitstring probabilities versus system size $N$. (b) Energy versus $N$, which ideally vanishes. (c) Observed post-selection success probability versus $N$.}
\label{fig:xi-state-m2}
\end{figure*}

Results obtained from the NISQ hardware are typically prone to multiple sources of error, leading to deviations from ideal unitary calculations. Here we quantify these errors by estimating two figures of merit: first, the Bhattacharyya distance 
\be
\label{eq:bd}
D(p, q) = - \ln \sum_{x} \sqrt{p(x) q(x)}
\ee
that captures the divergence between the exact and the observed distributions, $p(x)$ and $q(x)$, of the measured bitstrings $x$. Second, we use the expectation value of the parent Hamiltonian of the state $|\xi\rangle$, namely~\cite{Iadecola20}
\be
H_\xi = \sum_{i=2}^{N-1} P_{i-1}
\lf[\xi^{-1} P'_{i} + \xi P_{i} - (-1)^{i} X_{i} \rt]
P_{i+1},
\label{eq:H_xi}
\ee
that vanishes $\langle  H_\xi\rangle = 0$ in the case of ideal noiseless preparation of the state $|\xi\rangle$~\footnote{We note that $\ket{\xi}$ is the unique zero-energy eigenvector of $H_\xi$ in the Fibonacci Hilbert space, and that the spectrum of $H_{\xi}$ is positive semi-definite.}. We collect 10 observed samples of these quantities, along with the success probability of the post-selection protocol, for a range of different system sizes $N$ and block sizes $m$. Each sample measurement is made with $10^4$ shots for the energy and the success probability, and $10^5$ shots for the Bhattacharyya distance. The individual samples are depicted as round dots in \cref{fig:xi-state-n14}~and~\ref{fig:xi-state-m2}, and their average values are connected with solid lines. See the captions of \cref{fig:xi-state-n14}~and~\ref{fig:xi-state-m2} for the specification of used device nodes.

When the qubits are measured immediately after state preparation, without performing further unitary operations for, e.g., time evolution, the two-step process of applying Toffoli gates and post-selecting on ancilla bits can be reduced to the classical post-selection of $N$-qubit output bitstrings. Such replacement reduces the room for error, since Toffoli ($\mathrm{CCNOT}$) is a non-native gate that must be compiled into multiple imperfect two-qubit gates (e.g., $\mathrm{CPHASE}$, $\mathrm{CZ}$, $\mathrm{XY}$) before running on QPUs~\footnote{An alternative to decomposing a Toffoli gate into native two-qubit gates is to tune up a two-qutrit gate pulse that maps $|11\rangle$ to $|02\rangle$. The XY${}_{02}$ gate can serve as a building block for many doubly-controlled gates \cite{abrams2019_xy_qutrit}. See \cite{hill2021_rigetti_ccnot} for the implementation and benchmarking of three-qubit gates, e.g., Toffoli, with three-level control pulses.}.
It also ``unstitches" the whole circuit back into $\frac{N-2}{m}$ decoupled blocks of $\mathfrak{U}_{\xi}(m)$, which introduces certain computational advantages. For example, the asynchronous execution of fragmented circuits allows us to simulate systems larger than the hardware size, or to reduce the overall error by avoiding the usage of low-fidelity qubits, at an exponential overhead of classical post-processing \cite{Peng_2020,circuit-shadow}. With this in mind, we repeatedly use the same $m$-qubit sublattice that exhibit the best average readout and gate fidelities, and then classically combine the measurement outcomes with the Fibonacci constraint to simulate $N = \alpha m + 2$ qubit observables ($\alpha \in \mathbb{N}$).

\begin{figure*}
\centering
\includegraphics[width=0.485\textwidth]{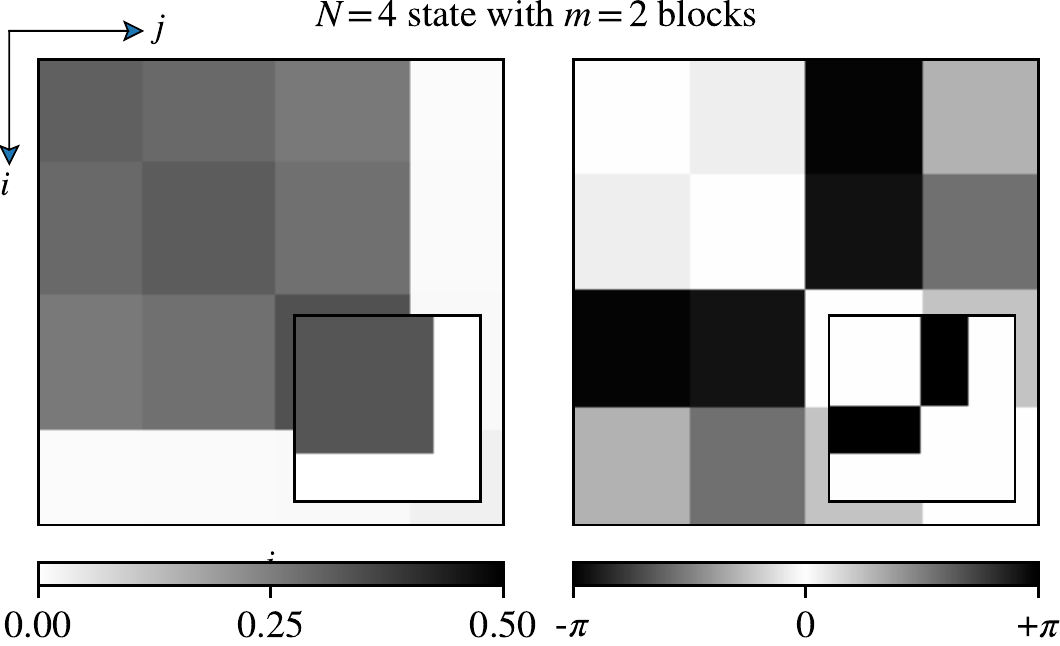}
\hfill
\includegraphics[width=0.485\textwidth]{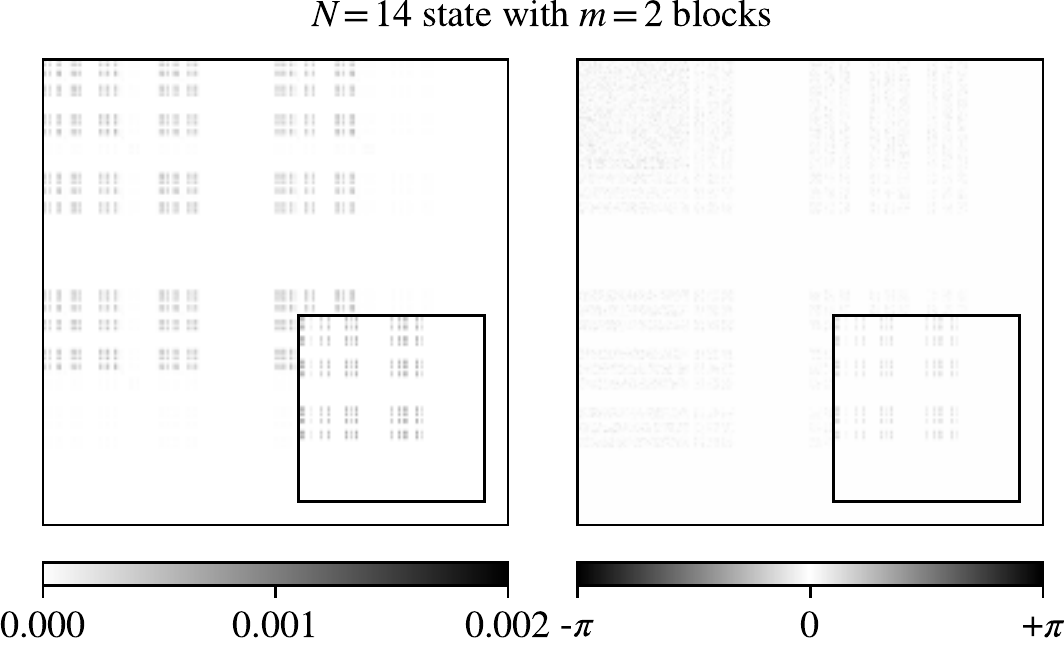}
\caption{State tomography of the $|\xi = 1\rangle$ states for $N=4$ and $14$, constructed from multiple copies of $\mathfrak{U}_{\xi}(m=2)$ with the classical post-processing. It is computed using $\{105, 104\}$ qubits on Aspen M-3. The $(i, j)$ block of the left / right density subplots shows $|| \rho_{i,j} || $ / $\text{arg}(\rho_{i,j})$, respectively, where $\rho_{i,j} \equiv \langle i | \xi = 1\rangle \langle \xi=1 | j \rangle$ and the integers $i,j$ are in the binary representation. The nearly empty density matrix plot in the right panel ($N=14$) illustrates the sparsity of the state $|\xi\rangle$. The inset plots show the ideal density matrices for comparison.}
\label{fig:xi-state-tomo}
\end{figure*}

\cref{fig:xi-state-n14} displays the results of the $N=14$ state preparation experiment with different block sizes $m$, to which the circuit $\mathfrak{U}_{\xi}(m)$ is applied. We first discuss the raw values without error mitigation, shown as green dots. As $m$ increases, both the energy expectation value $\langle H_\xi \rangle$ and the Bhattacharyya distance deviate farther from their ideal values of $0$. At the same time, the success probability grows with increasing $m$. These manifest the trade-off between the error and the post-selection success probability: the use of a greater number of smaller circuit fragments facilitates higher-fidelity execution of the circuit, while demanding an increased sample complexity. Note that our experimental protocol is a special case of quantum divide-and-conquer~\cite{Peng_2020, circuit-shadow}, where each block does not depend on another block's measurement result.

\cref{fig:xi-state-m2} summarizes the output of the $4 \leq N \leq 14$ state preparation circuit that merges $\frac{N-2}{2}$ copies of $\mathfrak{U}_{\xi}(m=2)$. 
As the system size $N$ increases, the Bhattacharyya distance stays close to $0$. While the energy deviation shows an apparent $O(N)$ scaling, its values are considerably less than those with larger $m$ building blocks.
We also illustrate in \cref{fig:xi-state-tomo} the state tomography of $|\xi=1\rangle\langle\xi=1|$ in the computational basis. In particular, its right panel highlights the sparsity of $|\xi\rangle$ states at $N=14$.

Finally, we attempt to improve the precision of experimental results by adopting error mitigation techniques, which produce the blue and orange dots in \cref{fig:xi-state-n14}~and~\ref{fig:xi-state-m2}. The estimated expectations usually deviate from true underlying values due to readout errors. One straightforward approach to mitigate the impact of readout errors is to calibrate the confusion matrix, 
\begin{align}
    M_{ij} \equiv \text{Pr}(\text{observed bits }i\,|\,\text{prepared bits }j)
\end{align}
and estimate the correct measurement bits that would be obtained in the absence of readout errors by applying the inverse of the confusion matrix. A more refined version of this approach is to replace the direct matrix inversion with the constrained optimization, called the iterative Bayesian unfolding \cite{Nachman2020_readout_unfolding}. In our QPU experiments, we apply this technique by calibrating the confusion matrices for each qubit group associated with the octagonal sublattice in the Aspen-M device layout.
% To handle the measurement error, we apply the Bayesian unfolding \cite{Nachman2020_readout_unfolding} that uses the confusion matrix of 8-bit strings associated with Aspen-M's octagonal layout. 

The samples obtained from ``corrected" bitstrings are colored in blue. 
They amplify the error measured in the Bhattacharyya distance and underestimate $\langle H_\xi \rangle$ when compared to the raw samples. 

We note that, for the $\mathfrak{U}_{\xi}(m=2)$ block calculation, most blue samples have negative energy that violates the positive semi-definiteness of the Hamiltonian Eq. (\ref{eq:H_xi}). 
The presence of the spurious samples is apparently an artifact of the Bayesian unfolding. It likely occurs due to a phase error since the sample Bhattacharyya distances are close to $0$. When using larger circuit blocks $\mathfrak{U}_{\xi}(m>2)$, however, the Bayesian unfolding, combined with randomized compilation and readout symmetrization, helps to lower the estimated energy $\langle H_\xi \rangle$ while maintaining the Bhattacharyya distance near the same level. 

Readout symmetrization is a technique that mitigates the asymmetric effects of noise \cite{Smith2021_readout_symm}. It compiles multiple symmetrized programs for each original measurement. These compiled programs resemble the original program, except for a quantum bit-flip performed just before measuring the qubit and a subsequent classical bit-flip of the measurement result. The outcomes of these programs are then combined to generate a set of `symmetrized' readout results. Randomized compiling also adopts a similar approach by compiling multiple logically identical circuits, in which gates are randomized through virtual conjugation with Pauli operators \cite{Wallman2016_randomized_compiling}. By aggregating the outcomes of these equivalent unitary programs, this technique enhances circuit execution performance by suppressing coherent errors, which grow quadratically with circuit depth, and instead converting them into stochastic errors that accumulate linearly. The measurement results obtained from both circuit compilation techniques, using $30$ logically equivalent circuits, are presented as individual orange dots.

Moreover, we implemented the full state preparation protocol on an IBM QPU (``ibmq\_montreal") without using fragmented circuits to reduce the circuit size. The IBM-QPU results show that the protocol's performance increases with decreasing block size $m$ and system size $N$, consistent with the abovementioned Rigetti-QPU results. More details of the IBM-QPU experiments are provided in \cref{sec:ibm_result}.
We have also tried to simulate the system dynamics with the prepared state being the initial state, but the coherence properties of the QPU limit this effort.

\section{Preparing the States $\ket{\mathcal S_k}$}
\label{sec: Skprep}

Since generic states $\ket{\mathcal S_k}$ in this tower have entanglement scaling logarithmically with system size (see Sec.~\ref{sec:model}), they cannot be prepared with quantum circuits of constant depth. In this section, we demonstrate two polynomial-depth algorithms for preparing these states. The first, described in Sec.~\ref{sec:MPS}, relies on building an MPS representation of the states $\ket{\mathcal S_k}$ and then converting this representation to a quasi-polynomial depth quantum circuit. The second, described in Sec.~\ref{sec:variational}, is a variational strategy that uses a polynomial-depth variational ansatz circuit to represent the states. Finally, we provide a proof-of-principle realization of these states on quantum hardware in Sec.~\ref{sec:Sk_QPU}, where we also write down a simplified linear-depth circuit for preparing the highest-weight state in the tower, $\ket{\mathcal S_{N/2-1}}$.

\subsection{Quasi-Polynomial Depth Circuit from Matrix Product State Representation}
\label{sec:MPS}

To facilitate the discussion below, we rewrite the tower of scarred eigenstates for $N$ sites as follows [see Eqs.~\eqref{eq:Sk}, \eqref{eq:Qdag}, and \eqref{eq:tilde}]:
\begin{align}
\label{eq:Ddef}
    |\tilde{\mathcal S}_k\rangle= \ket 0 \otimes |\mathcal D^{N-2}_k\rangle\otimes \ket 0,
\end{align}
where $\ket{\mathcal D^{N-2}_k}$ is an equal-amplitude superposition of all bitstrings of length $N-2$ with Hamming weight $k$ (i.e., containing $k$ $1$s) that obey the Fibonacci constraint. The state $\ket{\mathcal D^m_k}\propto \mathcal P_{\rm fib}\ket{d^m_k}$, where the Dicke state $\ket{d^m_k}$ is the equal-amplitude superposition of \textit{all} length-$m$ bitstrings with Hamming weight $k$. While Dicke states can be prepared in depth $O(mk)$ using the recursive strategy proposed in Ref.~\cite{Bartschi19}, we have not found an analogous strategy to prepare the projected Dicke states $\ket{\mathcal D^m_k}$ (see Sec.~\ref{sec: conclusion} for further discussion on this point). We therefore resort to an alternative approach that prepares the desired states in depth $O(mk \log^2 k)$, which is identical to the Dicke state preparation circuit up to an asymptotically minor multiplicative factor of $\log^2 k$. For generic scarred eigenstates $\ket{\mathcal S_k}$, which are written in terms of $\ket{\mathcal D^m_k}$ with $m=N-2$ and $k=O(N)$, this translates to a circuit with depth $O(N^2\log^2 N)$. However, for states in the tails of the tower (i.e., ones for which $k$ is finite), the resulting circuits are of linear depth. 

Standard algorithms exist for converting an MPS with $m$ sites and bond dimension $\chi$ into $m$ unitaries of width $\ceil{\log_{2}{2\chi}}$~\cite{Huggins_2019, PhysRevA.101.032310}. We first describe how to represent $\ket{\mathcal{D}^{m}_{k}}$ as an MPS, and then convert this MPS to a unitary circuit for state preparation. 

One can express a matrix product state on $m$ sites with open boundary conditions as 
\begin{equation}\label{eq:mps}
    \sket{\psi} = \sum_{\vec{s}}\bra{L}M^{s_{1}}M^{s_{2}}\ldots M^{s_{m}}\ket{R}\ket{\vec{s}},
\end{equation}
where $\ket{\vec s}=\bigotimes^m_{i=1}\ket{s_i}$ are computational basis states labeled by bitstrings, $M^{s_{i}}$ are square matrices of size $\chi$, and $\bra{L}$ and $\ket{R}$ are respectively $\chi$-dimensional row and column vectors implementing open boundary conditions. One can view an MPS as a representation of a deterministic finite automaton (DFA), where the matrices $M^{s_i}$ correspond to a transition matrix $M$ for the DFA states~\cite{PhysRevA.78.012356}. Since, for any $i$, $M^{s_i}$ is a square matrix of size $\chi$, this DFA will have $\chi$ states, and $M^{s_{i}}_{j,k}$ will be nonzero if the character $s_{i}$ takes the state $j$ to the state $k$. The bra $\bra{L}$ and ket $\ket{R}$ denote the initial and final states of the DFA, respectively. It is important to note that the states in the DFA live in the auxiliary bond space of the MPS and are not quantum states themselves. 

\begin{figure}[t]
    \centering
\includegraphics[width=0.5\linewidth]{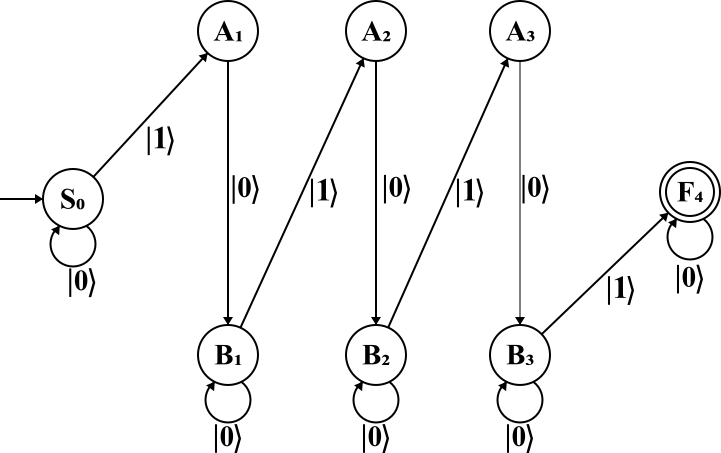}
    \caption{A graphical representation of the DFA transition matrix \eqref{eq:DFATransition} for the state $\sket{\mathcal D^*_4}$.}
    \label{fig:proj_dicke_dfa}
\end{figure}

DFAs are used to represent regular expressions, or string expressions using a finite set of characters. To see how we can use this language to represent quantum states, consider $\ket{\mathcal{D}^{4}_{2}}$:
\begin{equation}
    \ket{\mathcal{D}^{4}_{2}} = \frac{1}{\sqrt{3}}\left(\ket{0101} + \ket{1001} + \ket{1010}\right)
\end{equation}
The set of computational basis states in the superposition forms a language $\{\ket{0101}, \ket{1001}, \ket{1010}\}$ from which a regular expression can be constructed, where the characters in the regular expression are spin states $\{\ket{0}, \ket{1}\}$. For the states $\ket{\mathcal{D}^{*}_{k}} = \sum_{n=0}^{\infty}\ket{\mathcal{D}^{n}_{k}}$, the corresponding regular expression is 
\begin{align}
\sket{0}^{*}\left[\sket{1} \sket{0} \sket{0}^{*}\right]^{k-1}\sket{1}\sket{0}^{*},
\end{align}
where $^{*}$ is a Kleene star $(a^{*} = I + a + aa + aaa + ...$ where $I$ is identity (or empty string)$)$ and is only applied to $\ket{0}$, and $\left[.\right]^{m}$ means to repeat the term inside the bracket $m$ times.

An example of a DFA representing the regular expression for $\ket{\mathcal{D}^{*}_{k}}$ (with $k=4$) is shown in Fig.~\ref{fig:proj_dicke_dfa}. 
Note that if we limit the number of transitions through the DFA to $m$, the regular expression for $\ket{\mathcal D^*_k}$ reduces to that for $\ket{\mathcal{D}^{m}_{k}}$. For general $k$, the DFA is defined on $2k$ states $\{\mathrm{S}_{0}, \mathrm{A}_{j}, \mathrm{B}_{j}, \mathrm{F}_{k}\}_{j=1\ldots k-1}$. The DFA starts at $\mathrm{S}_{0}$ and ends at $\mathrm{F}_{k}$.  The first computational state can either be $\ket{0}$, which transitions the DFA back to $\mathrm{S}_{0}$, or $\ket{1}$, which transitions the DFA to $\mathrm{A}_{1}$. We denote this by 
\begin{align}
\begin{split}
\mathrm{S}_{0}(\ket{0}) &\to \mathrm{S}_{0}\\
\mathrm{S}_{0}(\ket{1}) &\to \mathrm{A}_{1}
\end{split}
\end{align}
For the intermediate states, we have
\begin{align}
\begin{split}
\mathrm{A}_{j}(\ket{0}) &\to \mathrm{B}_{j}\\
\mathrm{B}_{j}(\ket{0}) &\to \mathrm{B}_{j}\\
\mathrm{B}_{j}(\ket{1}) &\to \mathrm{A}_{j+1}
\end{split}
\end{align}
Finally, the DFA terminates at the final state $\mathrm{F}_{k}$, where
\begin{align}
\begin{split}
\mathrm{B}_{k-1}(\ket{1}) &\to \mathrm{F}_{k}\\
\mathrm{F}_{k}(\ket{0}) &\to \mathrm{F}_{k}
\end{split}
\end{align}
All these transition rules can be summarized by a transition matrix $\mathcal{M}$, where $\mathcal{M}_{i,j}$ gives the character needed to take the DFA from state  $i$ to state $j$:
\begin{align}
\begin{split}
\mathcal{M}_{\mathrm{S}_{0},\mathrm{S}_{0}}   = \ket{0}, & \indent \mathcal{M}_{\mathrm{S}_{0},\mathrm{A}_{1}}   = \ket{1},  \\
\mathcal{M}_{\mathrm{A}_{j},\mathrm{B}_{j}}   = \ket{0}, & \indent \mathcal{M}_{\mathrm{B}_{j},\mathrm{A}_{j+1}} = \ket{1},  \indent  \mathcal{M}_{\mathrm{B}_{j},\mathrm{B}_{j}} = \ket{0},\\
\mathcal{M}_{\mathrm{B}_{k-1},\mathrm{F}_{k}} = \ket{1}, & \indent \mathcal{M}_{\mathrm{F}_{k},\mathrm{F}_{k}} = \ket{0}.  
\end{split}
\end{align}
The full transition matrix is then
\begin{widetext}
\begin{align}
\label{eq:DFATransition}
\mathcal{M}= \begin{blockarray}{cccccccc}
 \mathrm S_{0}    & \mathrm A_{1}    & \mathrm B_{1}    & \cdots & \mathrm A_{k-1} & \mathrm B_{k-1}  & \mathrm F_{k} & \\
 \begin{block}{(ccccccc)c}
 \\
\sket{0} & \sket{1} & 0        & \cdots & 0       & 0        & 0        &~~ \mathrm S_{0} \\
          0        & 0        &\sket{0} & \cdots & 0       & 0        & 0        & ~~\mathrm A_{1} \\
          0        & 0        & \sket{0} & \cdots & 0       & 0        & 0        & \mathrm ~~B_{1} \\
          \vdots   & \vdots   & \vdots   & \ddots & \vdots  & \vdots   & \vdots   & \vdots \\
          0        & 0        & 0        & \cdots & 0       & \sket{0} & 0        & ~~\mathrm A_{k-1} \\
          0        & 0        & 0        & \cdots & 0       & \sket{0} & \sket{1} & ~~\mathrm B_{k-1} \\
          0        & 0        & 0        & \cdots & 0       & 0        & \sket{0} & ~~\mathrm F_{k} \\\\
 \end{block}
\end{blockarray}
\end{align}
% \begin{align}
% \label{eq:DFATransition}
%     \mathcal{M}=\begin{pNiceMatrix}[first-row,last-col,
%                           code-for-first-row=\scriptstyle,
%                           code-for-last-col=\scriptstyle,
%                           columns-width=auto]
%           \mathrm S_{0}    & \mathrm A_{1}    & \mathrm B_{1}    & \cdots & \mathrm A_{k-1} & \mathrm B_{k-1}  & \mathrm F_{k} \\
%           \sket{0} & \sket{1} & 0        & \cdots & 0       & 0        & 0        & \mathrm S_{0} \\
%           0        & 0        & \sket{0} & \cdots & 0       & 0        & 0        & \mathrm A_{1} \\
%           0        & 0        & \sket{0} & \cdots & 0       & 0        & 0        & \mathrm B_{1} \\
%           \vdots   & \vdots   & \vdots   & \ddots & \vdots  & \vdots   & \vdots   & \vdots \\
%           0        & 0        & 0        & \cdots & 0       & \sket{0} & 0        & \mathrm A_{k-1} \\
%           0        & 0        & 0        & \cdots & 0       & \sket{0} & \sket{1} & \mathrm B_{k-1} \\
%           0        & 0        & 0        & \cdots & 0       & 0        & \sket{0} & \mathrm F_{k} \\
%       \end{pNiceMatrix}
% \end{align}
To obtain the matrices from Eq.~\eqref{eq:mps}, we separate this transition matrix as $\mathcal{M} = M^{0}\ket{0} + M^{1}\ket{1}$, where
\begin{equation}
        M^{0} =\begin{pmatrix}
          1 & 0 & 0 & 0 & 0 & \cdots & 0 & 0 & 0  \\
          0 & 0 & 1 & 0 & 0 & \cdots & 0 & 0 & 0  \\
          0 & 0 & 1 & 0 & 0 & \cdots & 0 & 0 & 0  \\
          0 & 0 & 0 & 0 & 1 & \cdots & 0 & 0 & 0  \\
          0 & 0 & 0 & 0 & 1 & \cdots & 0 & 0 & 0  \\
          \vdots & \vdots & \vdots & \vdots & \vdots & \ddots & \vdots & \vdots & \vdots  \\
          0 & 0 & 0 & 0 & 0 & \cdots & 0 & 1 & 0  \\
          0 & 0 & 0 & 0 & 0 & \cdots & 0 & 1 & 0  \\
          0 & 0 & 0 & 0 & 0 & \cdots & 0 & 0 & 1  \\
    \end{pmatrix},~~
        M^{1} =\begin{pmatrix}
          0 & 1 & 0 & 0 & 0 & \cdots & 0 & 0 & 0  \\
          0 & 0 & 0 & 0 & 0 & \cdots & 0 & 0 & 0  \\
          0 & 0 & 0 & 1 & 0 & \cdots & 0 & 0 & 0  \\
          0 & 0 & 0 & 0 & 0 & \cdots & 0 & 0 & 0  \\
          0 & 0 & 0 & 0 & 0 & \cdots & 0 & 0 & 0  \\
          \vdots & \vdots & \vdots & \vdots & \vdots & \ddots & \vdots & \vdots & \vdots  \\
          0 & 0 & 0 & 0 & 0 & \cdots & 0 & 0 & 0  \\
          0 & 0 & 0 & 0 & 0 & \cdots & 0 & 0 & 1  \\
          0 & 0 & 0 & 0 & 0 & \cdots & 0 & 0 & 0  \\
    \end{pmatrix}\\
\end{equation}
% \begin{equation}
%     \NiceMatrixOptions{code-for-first-row=\scriptstyle,code-for-last-row=\scriptstyle}
%         M^{0} &=\begin{pNiceMatrix}
%           1 & 0 & 0 & 0 & 0 & \cdots & 0 & 0 & 0  \\
%           0 & 0 & 1 & 0 & 0 & \cdots & 0 & 0 & 0  \\
%           0 & 0 & 1 & 0 & 0 & \cdots & 0 & 0 & 0  \\
%           0 & 0 & 0 & 0 & 1 & \cdots & 0 & 0 & 0  \\
%           0 & 0 & 0 & 0 & 1 & \cdots & 0 & 0 & 0  \\
%           \vdots & \vdots & \vdots & \vdots & \vdots & \ddots & \vdots & \vdots & \vdots  \\
%           0 & 0 & 0 & 0 & 0 & \cdots & 0 & 1 & 0  \\
%           0 & 0 & 0 & 0 & 0 & \cdots & 0 & 1 & 0  \\
%           0 & 0 & 0 & 0 & 0 & \cdots & 0 & 0 & 1  \\
%     \end{pNiceMatrix},\qquad

%     \NiceMatrixOptions{code-for-first-row=\scriptstyle,code-for-last-row=\scriptstyle}
%         M^{1} =\begin{pNiceMatrix}
%           0 & 1 & 0 & 0 & 0 & \cdots & 0 & 0 & 0  \\
%           0 & 0 & 0 & 0 & 0 & \cdots & 0 & 0 & 0  \\
%           0 & 0 & 0 & 1 & 0 & \cdots & 0 & 0 & 0  \\
%           0 & 0 & 0 & 0 & 0 & \cdots & 0 & 0 & 0  \\
%           0 & 0 & 0 & 0 & 0 & \cdots & 0 & 0 & 0  \\
%           \vdots & \vdots & \vdots & \vdots & \vdots & \ddots & \vdots & \vdots & \vdots  \\
%           0 & 0 & 0 & 0 & 0 & \cdots & 0 & 0 & 0  \\
%           0 & 0 & 0 & 0 & 0 & \cdots & 0 & 0 & 1  \\
%           0 & 0 & 0 & 0 & 0 & \cdots & 0 & 0 & 0  \\
%     \end{pNiceMatrix}\\
% \end{equation}
\end{widetext}
Alternatively, $M^{0}$ is obtained by applying $\bra{0}$ to every entry in $\mathcal{M}$, and $M^{1}$ is obtained by applying $\bra{1}$ to every entry in $\mathcal{M}$. These are $2k\times2k$ matrices, and so the overall bond dimension of the MPS will be $2k$. Also, since we start at state $\mathrm{S}_{0}$ and end at state $\mathrm{F}_{k}$, the boundary vectors become
\begin{align}
\begin{split}
      \bra{L} &= \begin{pmatrix}1 & 0 & \cdots & 0 \end{pmatrix}\\
      \ket{R} &= \begin{pmatrix} 0 & \cdots& 0 & 1 \end{pmatrix}^{T}
\end{split}
\end{align}
One can now read Eq.~\eqref{eq:mps} from left to right as a traversal through the DFA in Fig.~\ref{fig:proj_dicke_dfa}, starting at DFA state $\mathrm{S}_{0}$, making $m$ transitions according to the spin configuration $\vec{s}$, and ending at DFA state $\mathrm{F}_{k}$.

\begin{figure*}
    \centering
    \includegraphics[width=.9\linewidth]{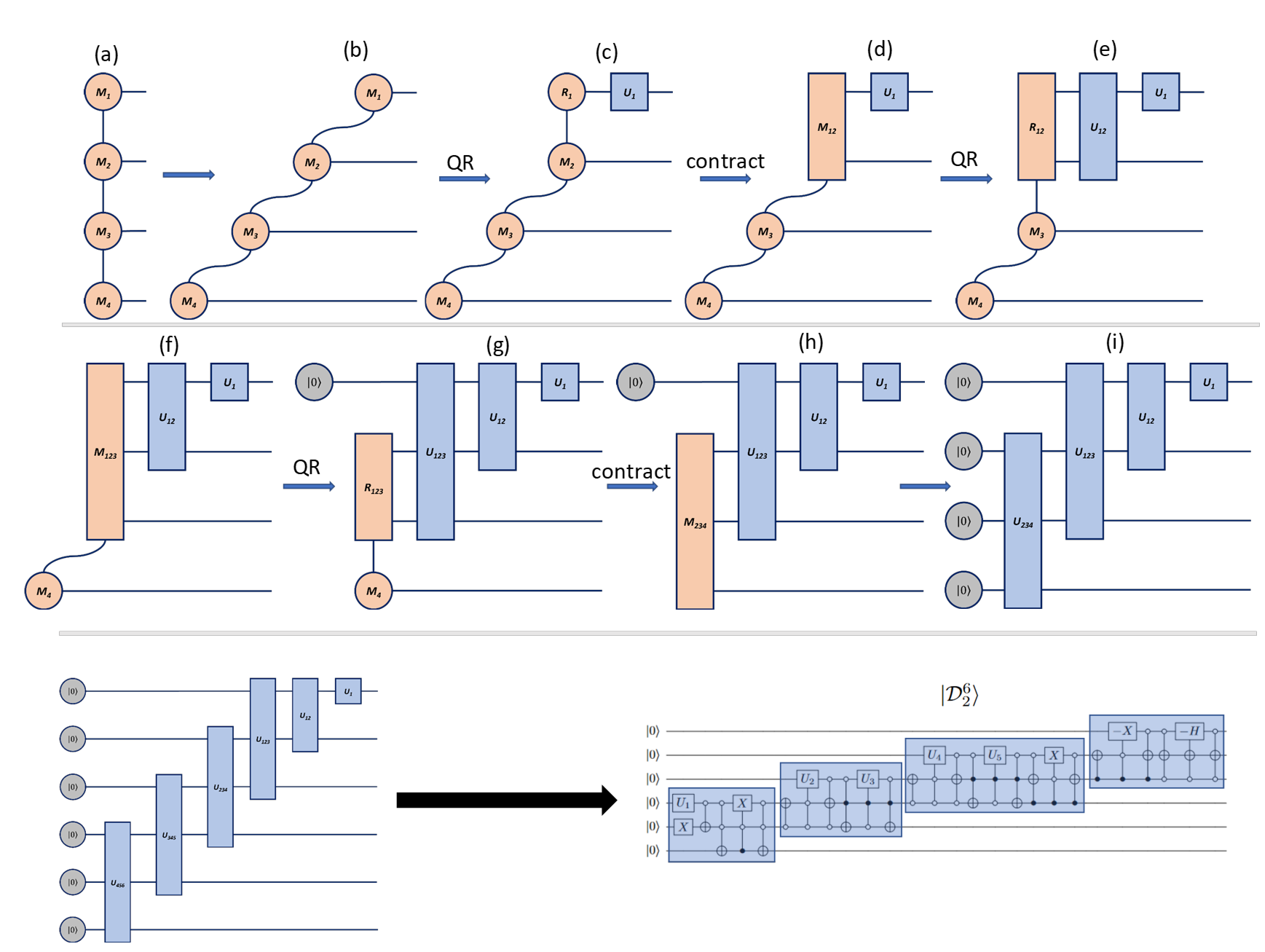}
    \caption{(top) Sequential generation of an MPS (a) into a quantum circuit (i) of 4 sites. The MPS is first ordered as shown in (b). A QR decomposition is performed on the top site giving a unitary $U_{1}$ and an upper triangular matrix $R_{1}$ resulting in (c). $R_{1}$ is then contracted with $M_{2}$ resulting in (d). The series of QR decompositions and contractions are performed as shown (e)-(h) until we each the final MPS circuit (i). Circuit diagram for the $\ket{\mathcal D^m_{k}}$ state with $m = 6$ and $k = 2$ obtained from the MPS representation. (bottom) Once the unitaries are formed, we use Gray codes to decompose the unitaries into quantum gates.}
    \label{fig:mps_to_circuit2}
\end{figure*}

The resulting MPS can be converted into a quantum circuit by following the sequential preparation procedure of Ref.~\cite{Huggins_2019}. Figure~\ref{fig:mps_to_circuit2} illustrates how this is done through a series of QR decompositions and tensor contractions. Specifically, each node in the MPS $M^{(s_{i})}_{jk}$ is a rank-3 tensor, except those at the boundaries which are rank-2 tensors. To perform a QR decomposition, we group together a physical index of $M$ with a virtual bond index, denoted as $M_{(s_{i}~j),k}$. A QR decomposition will decompose this $d\chi_r \times \chi_l$ matrix as $M_{(s_{i}~j),k} = U_{(s_{i}~j),l} R_{l,k}$ where $U_{(s_{i}~j),l}$ is an $d\chi_l \times d\chi_l$ unitary matrix, and $R_{l,k}$ is a $d\chi_l \times \chi_r$ upper triangular matrix, where $\chi_l$ and $\chi_r$ are the left and right bond dimension of $M^{(s_{i})}_{jk}$. 

Figure~\ref{fig:mps_to_circuit2} depicts a 4-site MPS with $\chi=4$ and $d=2$. The top-most qubit has $\chi_l = 1$ and $\chi_r=4$, so we decompose $M_{1} = U_{1} R_{1}$, where $U_{1}$ is a $2 \times 2$ unitary, and $R_{1}$ is a $2\times 4$ upper triangular matrix. $R_1$ is then contracted with the rank-3 tensor $M_2$ to form the rank-3 tensor $(M_{12})_{(ij)k}$. We then perform another QR decomposition $M_{12} = U_{12} R_{12}$, where both $U_{12}$ and $R_{12}$ are $4 \times 4$, and then contract $R_{12}$ with $M_{3}$ to form the rank-4 tensor $M_{123}$. The $8\times 4$ matrix $M_{123}$ allows us to perform a thin QR factorization, where we can decompose $M_{123} = V_{123} R_{123}$, where $V_{123}$ is an $8\times 4$ isometry $(V^\dagger_{123}V_{123} = I)$ and $R_{123}$ is a $4\times4$ matrix. To write $V_{123}$ as a circuit, we embed this matrix into an $8\times 8$ unitary $U_{123}$ such that when we apply the state $\sket{0}$ to the extra degree of freedom in $U_{123}$, we get the isometry $V_{123}$. This process of QR factorization followed by contraction is performed until all MPS sites have been written as a unitary.

%An arbitrary n-qubit unitary can be prepared by a circuit containing $O(4^n)$ CNOT gates~\cite{1629135}. Orthogonalizing the MPS in one direction results in $m$ $\ceil{\log_{2}{4k}}$-qubit unitaries, so the total number of CNOTs needed to construct the entire circuit is $O(mk^2)$. Fig.~\ref{fig:qcmps} depicts an example preparation circuit for the state $\ket{\mathcal D^6_{2}}$. The circuit makes use of the following single-qubit unitaries:

\begin{figure*}[t!]
\centering
\includegraphics[width=.9\linewidth]{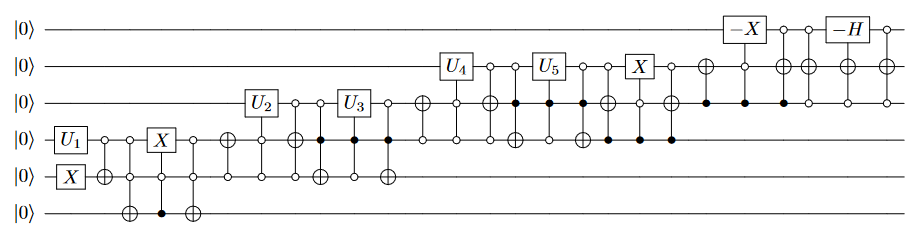}
\caption{Circuit diagram for the $\ket{\mathcal D^m_{k}}$ state with $m = 6$ and $k = 2$ obtained from the MPS representation. The unitaries $U_1$ through $U_5$ are listed in Eq~\eqref{eq:Us}}
\label{fig:qcmps}
\end{figure*}

Once this procedure is done, we are left with $m$ unitaries, each of which can be decomposed into $O(k)$ two-level unitaries. Using Gray codes, a two-level unitary, which here spans over  $O(\log k)$ qubits, can be decomposed into $O(\log k)$ Toffoli gates~\cite{NielsenChuang}. These  $O(\log k)$-qubit Toffoli gates require $O(\log k)$ CNOT gates~\cite{shende2008cnotcost}. The depth of the resulting circuit is then $O\left(mk\log^2k\right)$.  Fig.~\ref{fig:qcmps} depicts an example preparation circuit for the state $\ket{\mathcal D^6_{2}}$. The circuit makes use of the following single-qubit unitaries:

\begin{align}
\label{eq:Us}
\begin{split}
U_{1} &= \begin{pmatrix}
-\sqrt{\frac{2}{5}} & -\sqrt{\frac{3}{5}} \\
 -\sqrt{\frac{3}{5}}& \sqrt{\frac{2}{5}}
\end{pmatrix} \\
U_{2} & = \begin{pmatrix}
\sqrt{\frac{1}{2}} & \sqrt{\frac{1}{2}} \\
\sqrt{\frac{1}{2}}& -\sqrt{\frac{1}{2}}
\end{pmatrix}\\
U_{3} &= \begin{pmatrix}
-\sqrt{\frac{1}{4}} & \sqrt{\frac{3}{4}} \\
 \sqrt{\frac{3}{4}} & \sqrt{\frac{1}{4}}
\end{pmatrix}\\
U_{4} &= \begin{pmatrix}
\sqrt{\frac{2}{3}} &  \sqrt{\frac{1}{3}} \\
\sqrt{\frac{1}{3}} & -\sqrt{\frac{2}{3}}
\end{pmatrix}\\
U_{5} &= \begin{pmatrix}
-\sqrt{\frac{1}{3}} & \sqrt{\frac{2}{3}} \\
 \sqrt{\frac{2}{3}} & \sqrt{\frac{1}{3}}
\end{pmatrix}.
\end{split}
\end{align}

%% Abid: Add a summary discussing total CNOT count for the circuit in Fig. 10

\subsection{Polynomial Depth Variational Ansatz}
\label{sec:variational}

A strategy similar to the recursive linear circuit used in the preparation of the $|\xi\rangle$ state suggests a variational circuit architecture for creating the $|\mathcal S_k
\rangle$ state should exist. Up to unimportant relative phase factors, the $|\mathcal S_k
\rangle$ state is an equal-weight superposition of all bitstrings of $N$ sites, such that there are $k$ $1$s, the first and last sites are $0$, and there are no two neighboring $1$s. All such basis states can be generated by a four-qubit building-block unitary operator that transforms the state $|0100\rangle$ to a linear superposition of states $|0100\rangle$ and $|0010\rangle$~\cite{Yang20}, and acts trivially on all other 14 basis states.
\begin{table}
    \centering
    \begin{tabular}{ccccc}
    \hline \hline 
       $N$  & $k$ & $d$ & $n_a$ &$1-|\langle\psi|\mathcal S_k\rangle|^2$ \\
                                   \hline  
 13   & 5 &21 &20 & {$2.9\times 10^{-13}$}\\
        14   & 2 &55 &45 & {$4.9\times 10^{-5}$}\\
            14   & 3 &120 &41 & {$1.2\times 10^{-3}$}\\
14   & 4 &126 & 35& {$2.4\times 10^{-3}$}\\                  
        14   & 5 &56 & 27& {$2.2\times 10^{-13}$}\\
      
            14   & 6 &7 & 17& {$1.6\times 10^{-15}$}\\
                               
           ~~15~~ & ~~2~~ & ~66~& ~~52~~& ~~{$3.1\times 10^{-5}$}~~\\
                           
        15   & 3 &165 & 48& {$1.6\times 10^{-3}$}\\
                            
        15   & 4 &210 & 42& {$3.4\times 10^{-3}$}\\
        15   & 5 &126 & 34& {$2.1\times 10^{-3}$}\\

        15   & 6 &28 & 24& {$5.3\times 10^{-14}$}\\

        16   & 2 &78 & 60& {$1.7\times 10^{-5}$}\\
        16   & 3 &220 & 56& {$2.6\times 10^{-3}$}\\
           
        16   & 4 &330 & 50& {$4.4\times 10^{-3}$}\\
          
      16   & 5 &252 & 42& {$6.9\times 10^{-3}$}\\
          
      16   & 6 &84 & 32& $6.2\times10^{-13}$\\
       
      16   & 7 &8 & 20& $2.4\times 10^{-15}$\\
      \hline\hline  
    \end{tabular}
    \caption{The dimension of the constrained Hilbert space $d$, the number of variational angles $n_a$ and the minimum error $1-|\langle\psi|\mathcal S_k\rangle|^2$ achieved with 2000 local constrained optimization runs starting from random initial angles, for several $N$ and $k$.}
    \label{tab:var}
\end{table}
\begin{figure}[t]
\centering
\includegraphics[width=0.5\linewidth]{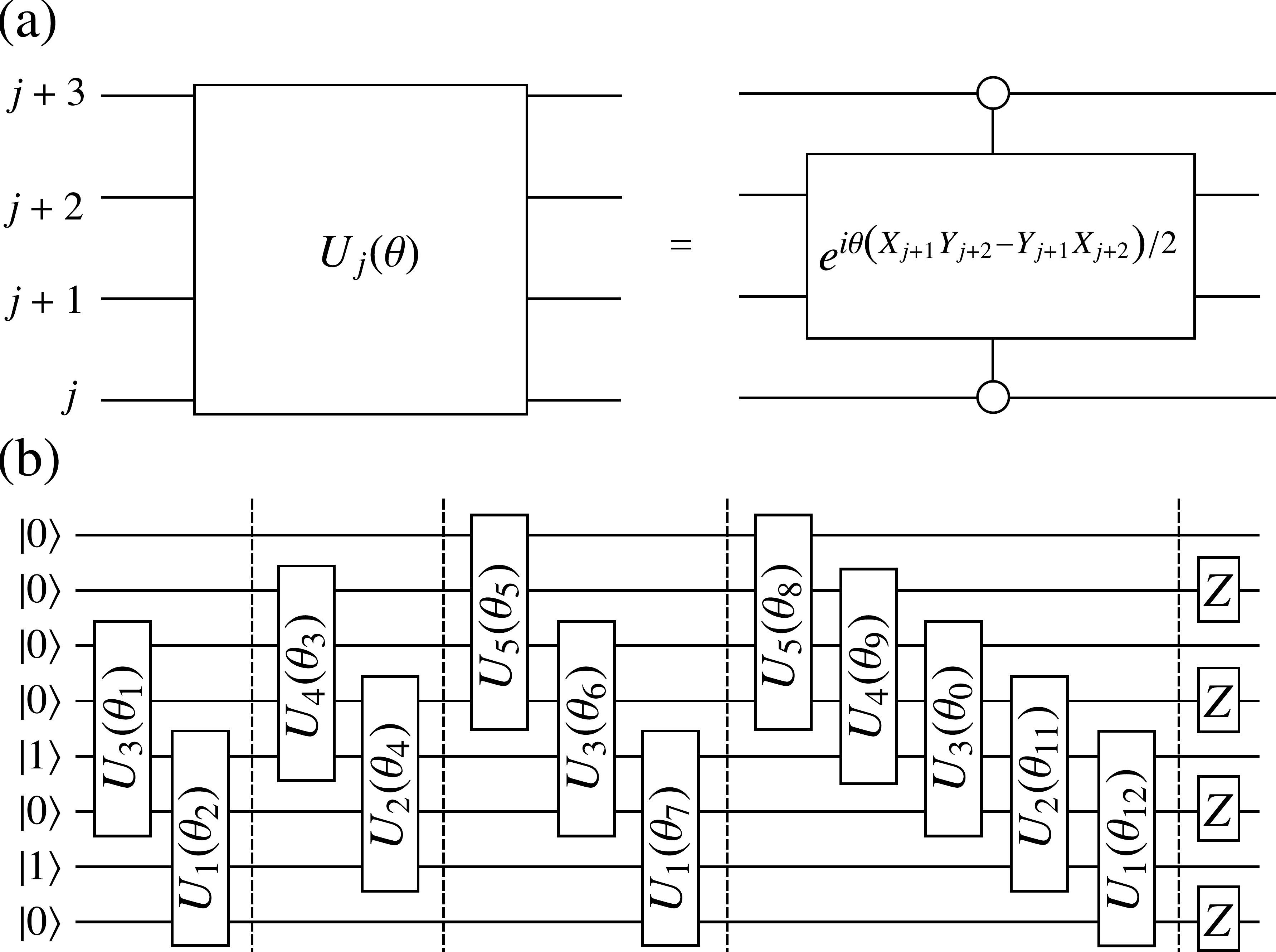}
\caption{(a) The four-qubit gate with one variational angle serves as a building block for constructing the variational ansatz. These gates conserve the number of ones and preserve the Fibonacci constraint while allowing transformations between all basis states in the constrained Hilbert space. (b) An example of the general architecture of the variational circuit for $N=8$ and $k=2$. }
\label{fig:var_circuit}
\end{figure}
One can construct the unitary by applying a two-qubit gate on the two middle sites controlled by the first and the last qubit such that the gates act nontrivially only when the first and the last qubits are 0, in which case the two middle qubits transform as
\begin{eqnarray*}
\ket{0}_{j+1}\!\ket{1}_{j+2}&\to &\cos(\theta_j)\ket{0}_{j+1}\!\ket{1}_{j+2}+ \sin(\theta_j)\ket{1}_{j+1}\!\ket{0}_{j+2}\\
\ket{1}_{j+1}\!\ket{0}_{j+2}&\to &\cos(\theta_j)\ket{1}_{j+1}\!\ket{0}_{j+2}- \sin(\theta_j)\ket{0}_{j+1}\!\ket{1}_{j+2}.
\end{eqnarray*}
We have chosen a building block with only real elements for easier optimization. 

Mathematically, we can write the unitary $U_j(\theta_j)$ acting on qubits $j$ to $j+3$ as as
\begin{equation}
\label{eq:Uj}
U_j(\theta)=\exp\left[i{\frac{\theta}{2}} P_j(X_{j+1}Y_{j+2}-Y_{j+1}X_{j+2})P_{j+3}\right],
\end{equation}
where the operators $P_j=\ket 0_j\bra 0_j$ implement the control on the first and last qubits and $(X_{j+1}Y_{j+2}-Y_{j+1}X_{j+2})$ acts as a Pauli-$Y$ on the $01$ and $10$ states of the two middle qubits, while annihilating $00$ and $11$. The building block $U_j$ is illustrated in Fig.~\ref{fig:var_circuit}(a).

We now construct an ansatz circuit.  Since the gates are designed to conserve both the total number of $1$s, i.e., $k$, and the Fibonacci constraint of no neighboring $1$s, we start with the following initial state that has $k$ $1$s: 
\begin{equation}
|\psi_0\rangle=|01\rangle^{\otimes k} |0\rangle^{\otimes N-2k}.
\end{equation}
In Fig.~\ref{fig:var_circuit}(b), we show an example of the circuit acting on the above initial state for $k=2$ and $N=8$. Our ansatz consists of multiple staircases of the $U_j$ gates. In the first layer, we apply $U_1...U_{2k-3}U_{2k-1}$ to move the last $1$ from qubit $2k$ to qubit $2k+1$ and generate all states with the last $1$ at or before qubit $2k+1$. The next layer, $U_2...U_{2k-1}U_{2k}$, generates all states with the last $1$ at or before $2k+2$. These staircases continue until we reach the staircase starting with $U_{N-3}$, acting on the last four qubits. All these staircases have unitaries acting on every other site. These layers generate all $d$ states. However, we have found that the number of variational parameters is insufficient to obtain an equal-weight superposition. Thus the ansatz also contains a complete staircase at the end with a unitary $U_1 U_2 ... U_{N-3}$ to equalize the probabilities. We then optimize the variational angles in these four-qubit unitaries to generate an equal-weight superposition of all 
$d$ basis states by minimizing $1-|\langle\psi|\mathcal S_k\rangle|^2$, where $|\psi\rangle$ is the state after the application of the variational unitaries. A final layer of $Z$ gates corrects the signs of the amplitudes.  

\begin{figure*}
	\centering
	\includegraphics[width=\linewidth]{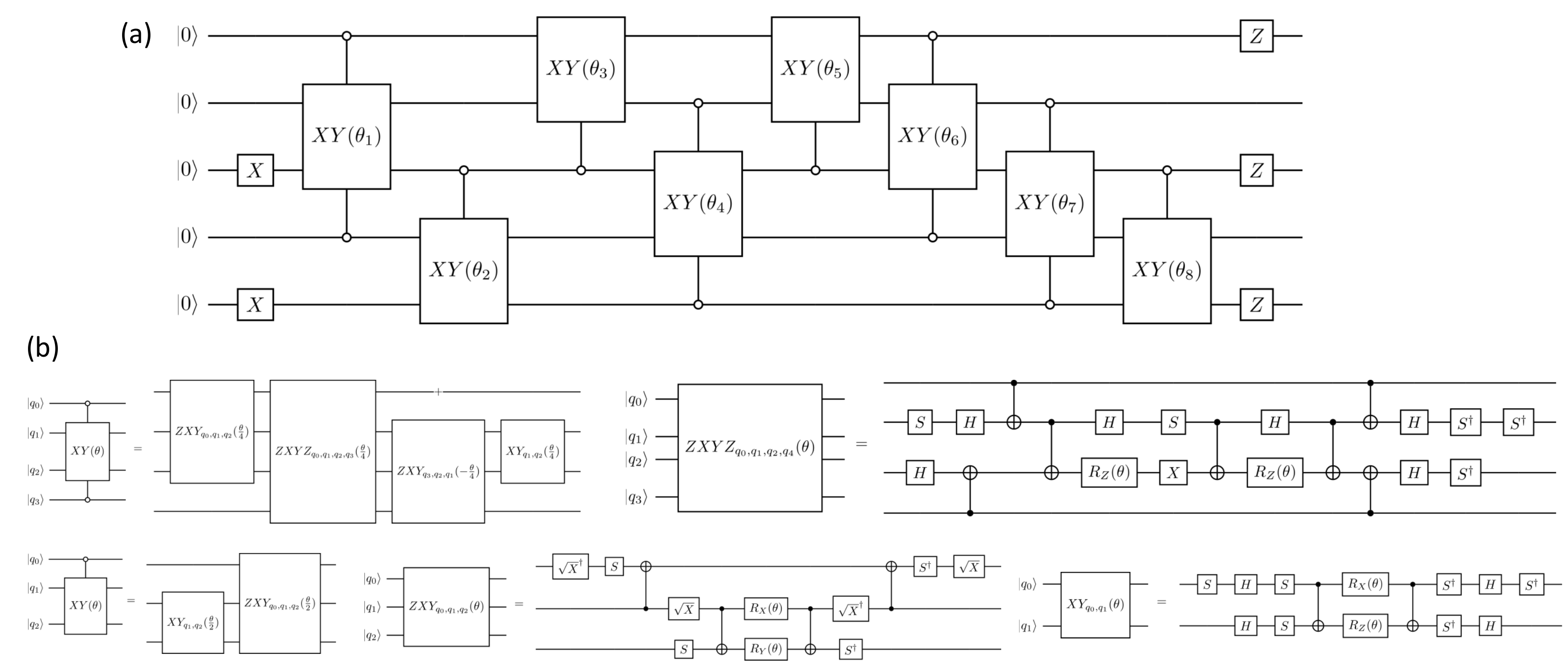}
	\caption{
		Variational circuit to prepare $\ket{\mathcal{D}^{m=5}_{k=2}}$. (a) The circuit consists of four single-controlled $XY$ and four double-controlled $XY$ gates, each parameterized by a single parameter. This circuit is pre-optimized by classical computers with machine-precision infidelity. (b) The controlled $XY$ gates are decomposed into CNOTs and single-qubit gates to be implemented on ``ibmq\_kolkata".
	}
	\label{fig:sk_expt_circuit}
\end{figure*}

Two important characteristics of this circuit are the dimension $d$ of the constrained Hilbert space and the number $n_a$ of variational parameters. While $d$ scales exponentially in the asymptotic limit, $n_a$ is quadratic in system size. The dimension of the constrained Hilbert space is $d=\mathcal N(N,k)=\binom{N-k-1}{k}$. We need our circuit to generate all the states subject to the Fibonacci constraint and the conserved number of $1$s. Thus, the basis states can be viewed as $k$ blocks of $01$ as partitions between $N-2k-1$ zeros with one zero tagged at the end of the chain. 
The number of variational angles $n_a$ is easy to calculate. For even $N$, the first and the second layer have $k$ gates each, the third and fourth layer $k+1$ gates, and so forth until we reach the staircase with $N/2-2$ gates containing every site from 2 to $N-4$. We then have a staircase with $N/2-1$ gates and then a full staircase with $N-3$ gates. Thus $n_a=2[k+(k+1)+...+N/2-2]+(N/2-1)+(N-3)$, which simplifies to
\begin{equation}
n_a=N^2/4-k(k-1)-2, \quad N {\rm ~even}.\end{equation}
Similarly for odd $N$, we have $n_a=2[k+(k+1)+...+(N+1)/2-2]+(N-3)$, which leads to
\begin{equation}
n_a=(N^2-1)/4-k(k-1)-2, \quad N {\rm ~odd}.\end{equation}
Importantly, the number of variational gates is quadratic in both $N$ and $k$.

The numerical optimization results are listed for several $N$ and $k$ in Table~\ref{tab:var}. The system exhibits many local minima (local minimization routines with different initial angles tend to converge to different optimal angles), making global optimization challenging. 
However, we have found that constrained local optimization with random initial angles between $0$ and $2\pi$ finds solutions with reasonably small error $1-|\langle\psi|\mathcal S_k\rangle|^2<0.01$. For many cases with smaller $n_a$ and $d$, and less complex optimization, the minimum error is zero to within machine precision, suggesting the $|\mathcal S_k\rangle$ state may be exactly reachable with the ansatz architecture. In practice, we can only find very good approximations for large $d$ and $n_a$. 

It would be interesting to determine whether the variational ansatz circuit proposed here can represent the $\ket{\mathcal S_k}$ states exactly. This would conclusively demonstrate that the scarred eigenstates of the model~\eqref{eq:H0} can be prepared in polynomial depth (as opposed to quasipolynomial depth as shown in Sec.~\ref{sec:MPS}), which is known to be possible for, e.g., Dicke states~\cite{Bartschi19}. Attempts to find an exact solution for the rotation angles in the circuit architecture proposed here yield $n_a$ coupled nonlinear equations that are intractable in all but the simplest cases. An interesting question for future work would be to investigate whether a modified circuit architecture involving gates of the form in Eq. \eqref{eq:Uj} can be used to obtain an analytically tractable system of equations for the rotation angles.

\subsection{QPU Results}
\label{sec:Sk_QPU}

\subsubsection{Pre-Optimized Variational Circuit}

In this subsection, we present a proof-of-principle experiment preparing $\ket{\mathcal S_k}$ using the variational circuit proposed in \cref{sec:variational} for one of the simplest cases, $N = 7$ and $k=2$. Similar to $\ket{\xi}$, we simulate only the middle five spins, as the two boundary spins are always fixed to $\ket{0}$ by the open-boundary conditions. In other words, we prepare $\ket{\mathcal D^{m=N-2}_k}$ defined in \cref{eq:Ddef}.

The five-qubit experiment involves the variational circuit shown in \cref{fig:sk_expt_circuit}(a). The variational parameters $\{ \theta_j\}_{j=1, 2, \cdots, 8}$ are pre-optimized on classical computers to obtain an infidelity of 0 to machine precision. The double-controlled $XY$ gates are equivalent to $U_j$ defined by \cref{eq:Uj}. The $U_j$ gates at the edge turn into single-controlled $XY$ gates since the two boundary spins are always in $\ket{0}$.
The single-controlled and double-controlled $XY$ gates are decomposed into CNOT and single-qubit gates as shown in \cref{fig:sk_expt_circuit}(b). With this decomposition, each single-controlled $XY$ gate is implemented by six CNOTs. The double-controlled $XY$ gate consists of 17 CNOTs, and it is further compiled down by Qiskit \cite{Qiskit} to 12 CNOTs. Hence, the five-qubit variational circuit consists of $72$ CNOT gates.

\begin{figure}
	\centering
	\includegraphics[width=0.5\linewidth]{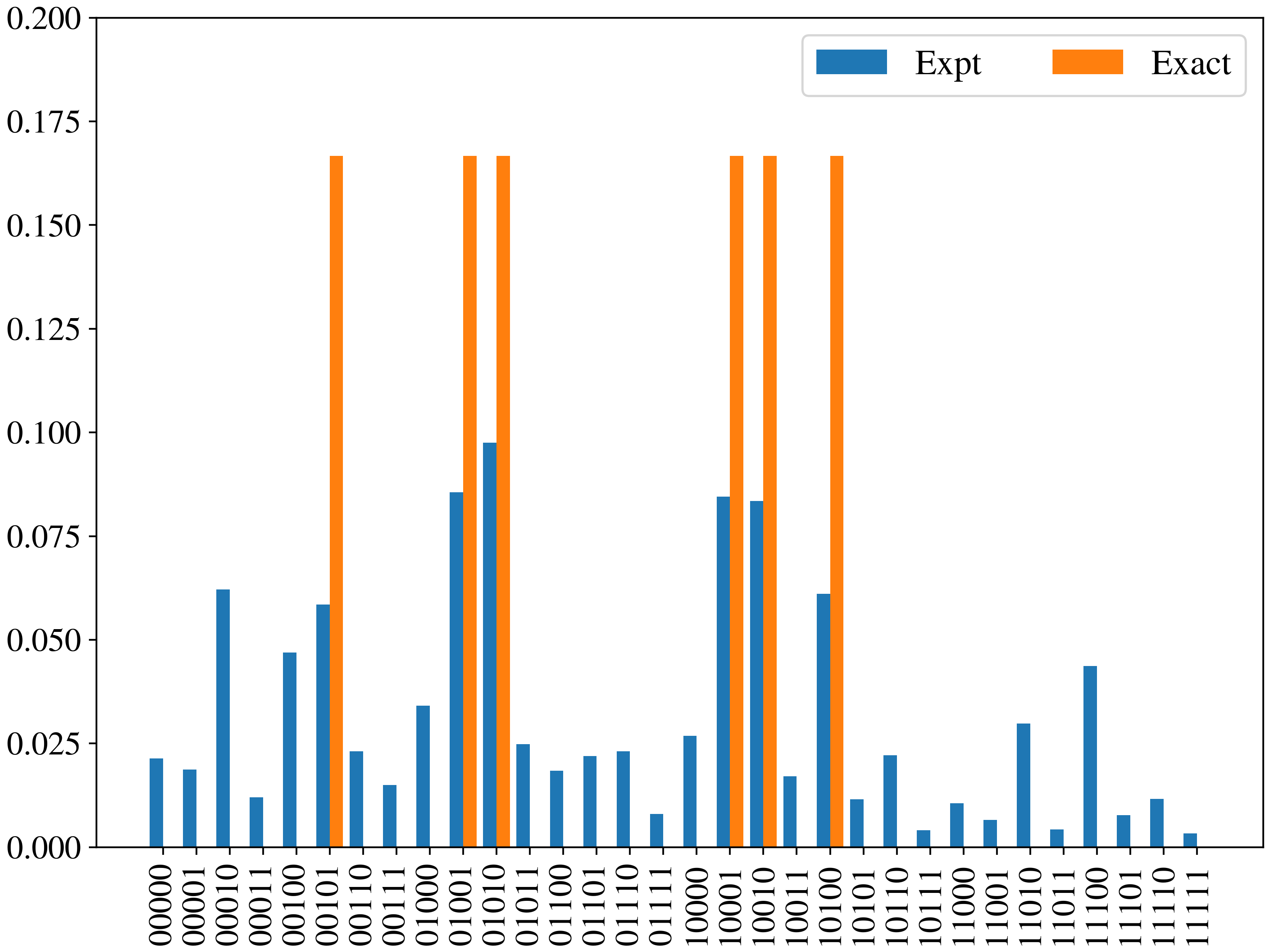}
	\caption{
		Probability distribution of $\ket{\mathcal{D}^{m=5}_{k=2}}$ prepared on ``ibmq\_kolkata". The state is prepared on qubits [14, 13, 12, 15, 18] and measured with 196608 shots. The errors are reduced by dynamical decoupling and readout error mitigation.
	}
	\label{fig:expt_variational_Sk}
\end{figure}

The pre-optimized circuit is experimentally simulated using an IBM QPU (``ibmq\_kolkata"). In order to reduce errors, the circuit is further optimized by inserting dynamical decoupling pulse sequences $\sqrt{X} \sqrt{X} \sqrt{X} \sqrt{X}$ to idling qubits, and readout error mitigation is performed on the measurement outcomes. The probability distribution of the five-qubit state is measured with 196608 shots and is shown in \cref{fig:expt_variational_Sk}.

The experimental result (blue bars) qualitatively matches the exact distribution (orange bars) with a few incorrectly high probabilities at 00010, 00100, 11010, and 11100. These incorrect bitstrings are one bit flip away from the correct bitstrings, and these errors are expected due to the high CNOT gate count of the variational circuit given the average CNOT infidelity of around 1\%. The Bhattacharyya distance between the experimental and exact distributions is 0.3809, which is comparable to the state-preparation result of $\ket{\xi}$ shown in \cref{sec:ibm_result}, albeit more qubits (up to 20 with ancilla qubits) are considered in that case.

Extending the experiment to larger $N$ on currently available QPUs is difficult due to the increased CNOT gate count exceeding the capacity of these QPUs. In order to perform the extended experiments on present-day hardware with high fidelity, it is necessary to carry out further simplifications like those discussed in the next subsection.

\subsubsection{Exact Circuit for $k=k_{\rm max}$}

\begin{figure}
\includegraphics[width=0.5\linewidth]{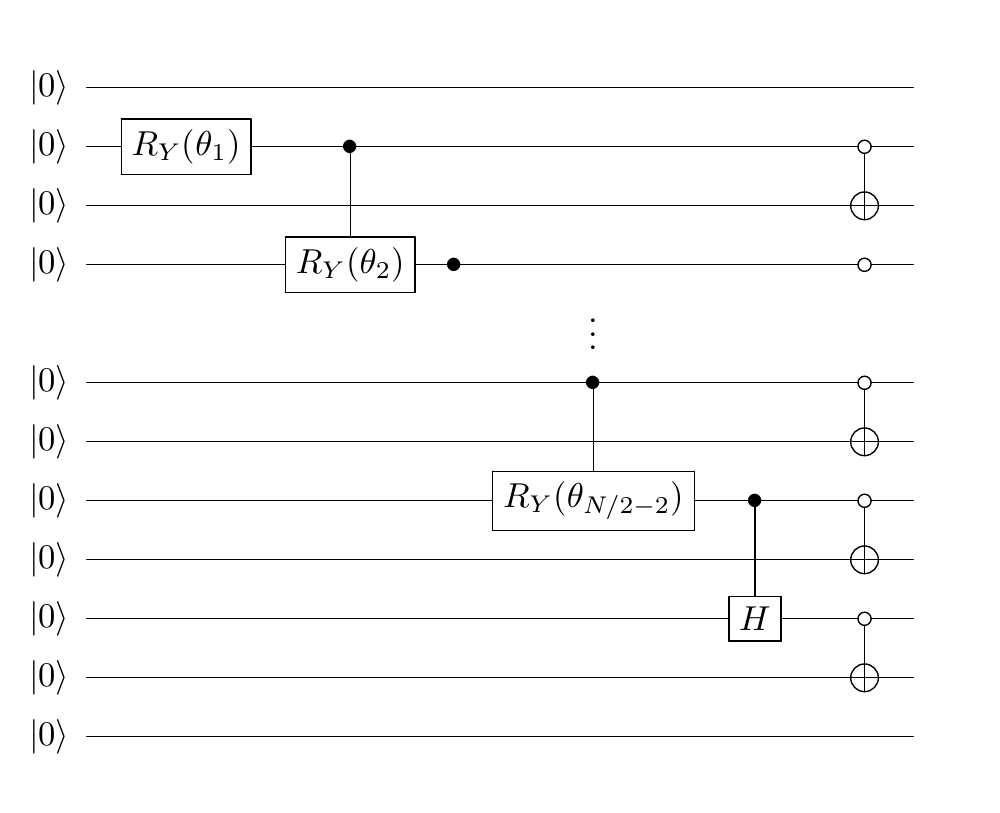}
% \begin{equation*}
% \Qcircuit @C=1em @R=0em @!R {
% \lstick{\ket{0}}& \qw & \qw & \qw & \qw&\qw&\qw&\qw & \qw \\
% \lstick{\ket{0}}& \gate{R_Y(\theta_1)} & \ctrl{2}& \qw &\qw &  \qw & \qw & \ctrlo{1} & \qw \\
% \lstick{\ket{0}}& \qw  & \qw& \qw & \qw & \qw & \qw &  \targ{}& \qw\\
% \lstick{\ket{0}}& \qw & \gate{R_Y(\theta_2)} & \ctrl{0}&\qw  &\qw & \qw & \ctrlo{0}& \qw\\
% &&&&\vdots&&&&\\
% \lstick{\ket{0}}& \qw & \qw & \qw & \ctrl{2} & \qw& \qw & \ctrlo{1} & \qw \\
% \lstick{\ket{0}}& \qw & \qw & \qw & \qw & \qw& \qw & \targ{} & \qw \\
% \lstick{\ket{0}}& \qw & \qw & \qw & \gate{R_Y(\theta_{N/2-2})} & \ctrl{2} & \qw & \ctrlo{1} & \qw \\
% \lstick{\ket{0}}& \qw &\qw & \qw &\qw & \qw & \qw & \targ{} & \qw\\
% \lstick{\ket{0}}& \qw &\qw & \qw &\qw &  \gate{H}&\qw & \ctrlo{1}& \qw\\
% \lstick{\ket{0}}& \qw &\qw & \qw &\qw & \qw & \qw & \targ{} & \qw\\
% \lstick{\ket{0}}& \qw &\qw & \qw &\qw & \qw & \qw & \qw& \qw
% }\end{equation*}
\caption{The $|\tilde{\mathcal{S}}_{k}\rangle$ state preparation circuit with $k=k_\text{max} \equiv \frac{N}{2}-1$. The rotation angles are $\theta_i = 2\tan^{-1}(\sqrt{N/2 - i})$. The scar state $\ket{\mathcal S_{\rm k_{\rm max}}}$ is obtained by applying $Z$ gates at the end of the circuit following Eqs.~\eqref{eq:tilde}.}
\label{fig:sk-state-prep-max-k}
\end{figure}

It is noticeable that the doubly-controlled gates are essential pieces in building up the state preparation circuits described in Secs.~\ref{sec:Linear}~and~\ref{sec:variational}. As we saw in the previous section, compiling them into the set of native gates often produces quantum programs relatively expensive in the two-qubit gate complexity. An important simplification occurs for the specific case of $k = k_\text{max} \equiv N/2 - 1$ for which an alternative circuit, shown in \cref{fig:sk-state-prep-max-k}, can be used that involves only $(N-3)$ two-qubit gates.

We can deterministically realize the maximal $k$ condition by setting the spin state of the $(i+1)$-th qubit to be opposite to that of the $i$-th qubit for all $i \in \{2, 4, \cdots, N-2\}$.
This can be implemented by initializing the target qubit in $|0\rangle$ and then acting with the controlled bit-flip operator,
\begin{align}
\mathrm{C}_{0}\mathrm{NOT}_{i, i+1} = X_{i} \cdot \mathrm{CNOT}_{i, i+1} \cdot  X_{i},
\label{eq:c0not}
\end{align}
triggered only when the control qubit is in the $|0\rangle$ state.
Next, the superposition state can be prepared with a variant of the unitary block Eq. (\ref{eq:linear_preparation_circuit}) where we modify the controlled rotations to be triggered only if the control qubit is in the $|1\rangle$ state.
Running it on the even-indexed qubits between $2\leq i \leq N-2$, we obtain a superposition of the following bitstring states,
\begin{align}
    |0(10)^a  (01)^b0\rangle \ \text{ satisfying } \ a+b = k_\text{max}.
\end{align}
Upon acting with Eq. (\ref{eq:c0not}) on all the adjacent $(i, i+1)$ pairs with $i \in \{2, 4, \cdots, N-2\}$, the superposition state becomes
\begin{align}
    \sum_{a+b = k_\text{max}} c_{a,b}\, |0(10)^a (01)^b0\rangle 
    \label{eq:circuit_state_sk}
\end{align}
where $c_{a,b} \in \mathbb{R}$ and $\sum_{a, b}c_{a,b}^2 = 1$. Lastly, we observe that the following Pauli rotation parameters,
\begin{align}
\theta_i = 2\tan^{-1}(\sqrt{N/2 - i}),
\label{eq:angle}
\end{align}
leads to the equal-amplitude superposition state $\ket 0 \otimes |\mathcal D^{N-2}_{k_\text{max}}\rangle\otimes \ket 0$,
since all coefficients become the same:
\begin{align}
c_{a,b} = \prod_{i=1}^{a} \sqrt{\frac{\frac{N}{2} - i}{\frac{N}{2} - i + 1}} \cdot \sqrt{\frac{1}{\frac{N}{2} - a}} = \sqrt{\frac{1}{N/2}}.
\end{align}
Figure~\ref{fig:sk-state-prep-max-k} illustrates the $|\tilde{\mathcal{S}}_{k_\text{max}}\rangle$ state preparation circuit.

We execute the state preparation circuit on Aspen-M-2 and Aspen-M-3.
Ideally, if we perform Pauli-$Z$ measurements after running the circuit, the resulting $N$-bitstrings must include $k_\text{max}$ non-consecutive $1$'s and begin and end with $0$ by construction. However, due to limitations of  current NISQ hardware, the circuit simulation of the $|\mathcal{S}_{k_\text{max}}\rangle$ state often induces an error in the total spin measurement $k = k_\text{max}$ and may violate the boundary condition or the Fibonacci constraint. Therefore, to increase the fidelity of the simulation, we post-select the measurement bitstrings along with optional application of the readout error mitigation and randomized compiling techniques.

\begin{figure*}
\centering
\includegraphics[width=\textwidth]{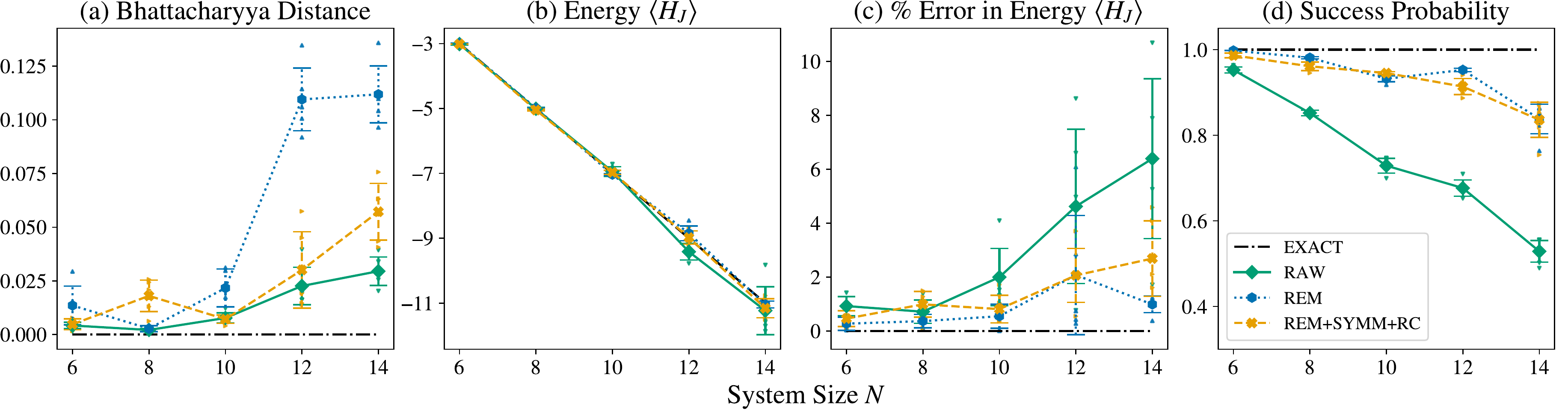}
\caption{
The execution results of the $|\mathcal{S}_{k_\text{max}}\rangle$ state preparation circuit on Aspen-M devices, which uses the $k_\text{max} = \frac{N}{2}-1$ qubit encoding for $6 \leq N \leq 14$. The indices of the participating Aspen nodes are: $\{147, 140\}$ (M-3) for $N=6$, $\{147, 140, 141\}$ (M-3) for $N=8$, $\{16, 17, 10, 11\}$ (M-2) for $N=10$, $\{11, 10, 17, 16, 1\}$ (M-2) for $N=12$, $\{26, 11, 10, 17,16, 1\}$ (M-2) for $N=14$.
We refer to the caption of Figure~\ref{fig:xi-state-n14} for a detailed explanation of the data labels (EXACT, RAW, REM, REM+SYMM+RC) and the number of repeated measurements.
We gather $5$ sample points for each quantity, with the value of $N$ varying, where the error bars denote the error of the mean across different samples.
(a) Bhattacharyya distance, which is a measure of the difference between the ideal and measured bitstring probabilities versus system size $N$. (b) Energy versus $N$, which ideally takes the value of $-(N-3)$. (c) Energy error versus $N$, computed from $100\, \frac{|\langle H_J \rangle + N-3|}{N-3}$. (d) Observed postselection success probability versus $N$.
}
\label{fig:sk-state-run}
\end{figure*}

Assuring the total spin selection rule allows us to introduce the compact encoding scheme that uses only $k_\text{max}$ qubits, i.e., indexed as $\{2, 4, \cdots, N-2\}$, to encode the $N$-qubit states satisfying $k = k_\text{max}$ without consecutive $1$'s. It is equivalent to decoupling the non-entangled boundary spins and replacing the error-prone quantum operator Eq.\ \eqref{eq:c0not} with deterministic post-processing. Its use of fewer qubits also brings the advantage of circumventing non-linear qubit connectivity and considerable error reduction. 
To make sure the Fibonacci constraint is satisfied, we  apply the following projection operator
\begin{align}
\mathcal{P} = \prod_{i=2,4, \cdots, N-4} (1-P_i P'_{i+2})
\label{eq:fibonacci-proj}
\end{align}
that removes $|01\rangle$ configurations in the compressed $k_\text{max}$-qubit encoding, or equivalently, $|0110\rangle$ configurations in the full $N$-qubit representation.

\begin{table*}
\centering
\begin{tabular}{cccccccc}\hline\hline
Before & After & Before & After & Before & After & Before & After \\\hline
$Z_{1}$ \text{and} $Z_{N}$ &  $1$ & $X_{2i}Z_{2i+1}$ & $0$ & $X_{2i}I_{2i+1}$ & $0$ & $X_{2i}X_{2i+1}$ & $X_{2i}$ \\ 
$Z_{2i}I_{2i+1}$ & $Z_{2i}$ & $Y_{2i}Z_{2i+1} $ & $0$ & $Y_{2i}I_{2i+1} $ & $0$ & $X_{2i}Y_{2i+1} $ & $-Y_{2i}$  \\
$I_{2i}Z_{2i+1}$ & $-Z_{2i}$ &   $Z_{2i}X_{2i+1}$ & $0$ & $I_{2i}X_{2i+1}$ & $0$ & $Y_{2i}X_{2i+1}$ & $Y_{2i}$   \\
$Z_{2i}Z_{2i+1}$ & $-1$ & $Z_{2i}Y_{2i+1}$ &$0$ & $I_{2i}Y_{2i+1}$ &$0$ & $Y_{2i}Y_{2i+1}$ &$X_{2i}$\\\hline\hline
\end{tabular}
\caption{Projection rule for $N$-qubit Pauli strings to Pauli operators on the $k_\text{max}$-qubit Hilbert space with $1\leq i \leq k_\text{max}$. 
}
\label{tbl:encoding}
\end{table*}

Having run the state preparation program, we measure the errors through the Bhattacharyya distance defined in Eq. \eqref{eq:bd} between the ideal and empirical bitstring distributions and the expectation value of the spin-$1/2$ Hamiltonian from Eq.~\eqref{eq:H0}. With the $k_\text{max}$-qubit encoding, the Hamiltonian is mapped to
\begin{align}
     H_0   =& \, \lambda  H_\lambda  + 
    \Delta  H_\Delta  + J  H_J 
\end{align}
where
\begin{align}
\begin{split}
     H_\lambda  &= 0, \quad H_\Delta  = 2, \quad \text{and}\\
     H_J  &=\,  Z_2 -  Z_{N-2} - \tfrac{N}{2} + 1 - \textstyle\sum^{N/2-2}_{i=1} Z_{2i}  Z_{2i+2}.
\end{split}
\label{eq:compressed_h0}
\end{align}
It is obtained by applying the encoding rules of Table~\ref{tbl:encoding}, where vanishing substitutions on the second and third columns impose the total spin selection rule $k = k_\text{max}$. This reduces the task of computing $\langle H_0 \rangle$ to measuring $\langle H_J \rangle$.

\cref{fig:sk-state-run} collects the results of the $|\mathcal{S}_{k_\text{max}}\rangle$  state preparation experiment, showing the Bhattacharyya distance, the energy $\langle H_J \rangle$ and its error, and the success probability of the Fibonacci projection, see Eq. \eqref{eq:fibonacci-proj}. 
Individual sample points are drawn as round dots, and their averages are connected as the solid line. Each dot aggregates $10^4$ shots for the energy and success probability, and $10^5$ shots for the Bhattacharyya distance. We record the qubit indices used for the QPU experiment in the caption of \cref{fig:sk-state-run}.

\begin{figure*}
\centering
\centering
\includegraphics[width=0.485\textwidth]{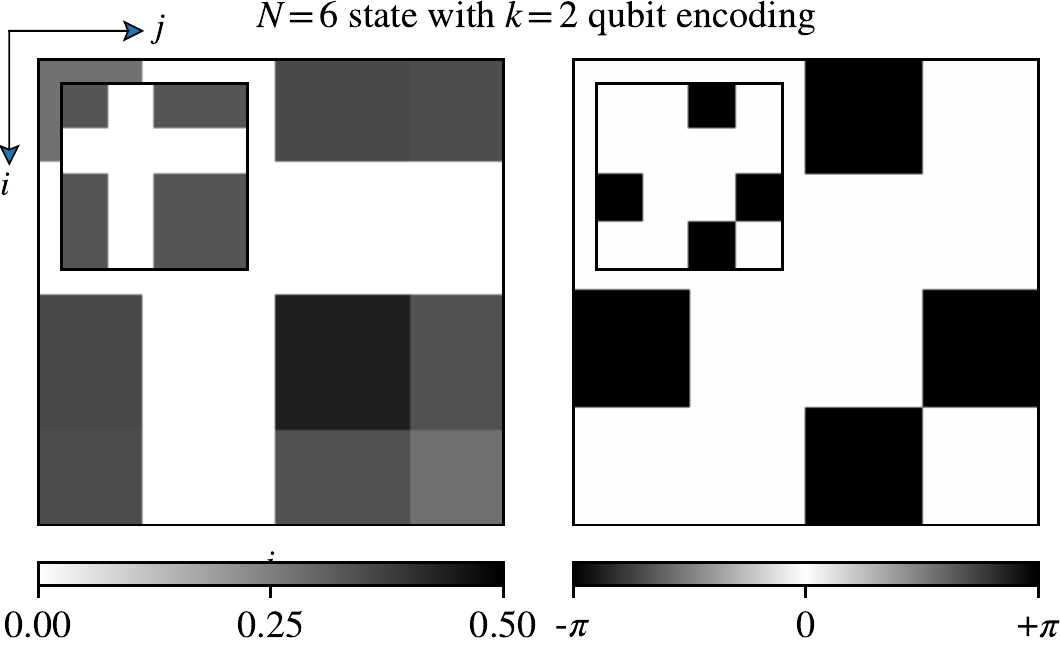}
\hfill
\includegraphics[width=0.485\textwidth]{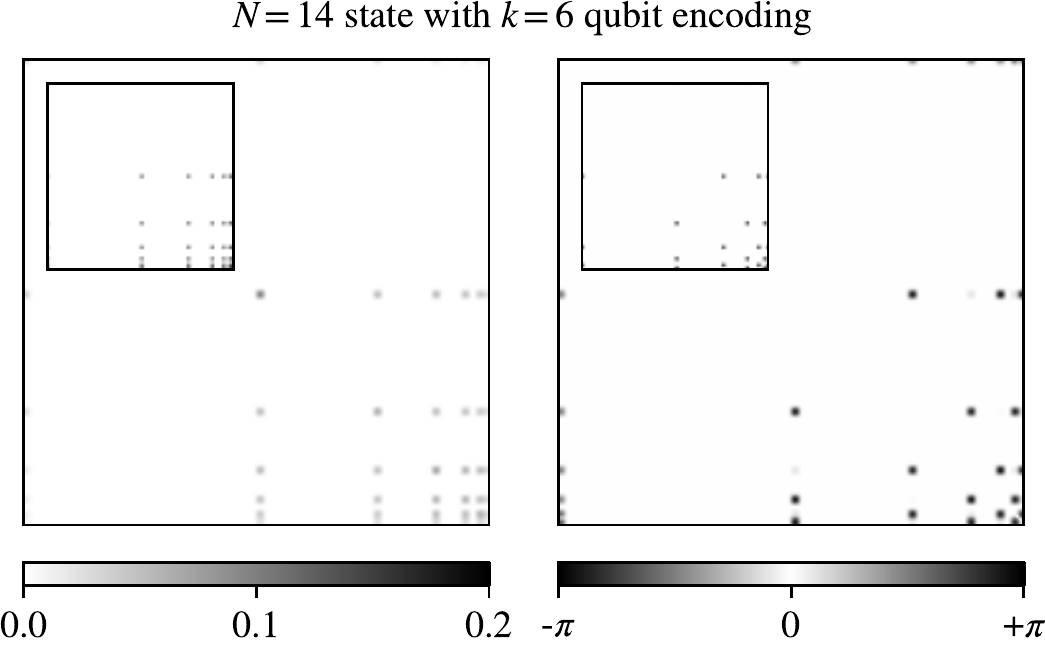}
\caption{The $|\mathcal{S}_{k_\text{max}}\rangle$ 
 state tomography in the compressed $k_\text{max}$-qubit encoding at $N=6$ and $14$. 
It uses the $\{147, 140\}$  ($N=6$) and $\{26, 11, 10, 17,16, 1\}$ ($N=14$) device qubits on Aspen M-3 and M-2, respectively. The $(i, j)$ block of the left and right density subplots shows $|| \rho_{i,j} || $ and $\text{arg}(\rho_{i,j})$, respectively, where $\rho_{i,j} \equiv \langle i |  \mathcal{S}_{k_\text{max}} \rangle \langle \mathcal{S}_{k_\text{max}} | j \rangle$ and the integers $i,j$ are in the binary representation. The nearly empty appearance of the density plot in the right panel ($N=14$) manifests the sparsity of the scar state $|\mathcal{S}_{k_\text{max}}\rangle$ which only occupies $N/2$ out of $2^{k_\text{max}}$ possible bitstrings.
The inset plots show the ideal density matrices for comparison.}
\label{fig:sk-state-tomography}
\end{figure*}

We consider the raw samples without the readout error mitigation and randomized compiling techniques, denoted as green dots. We note a rapid drop of the success probability as $N$ increases. Unlike the probabilistic $|\xi\rangle$ state preparation protocol in Section~\ref{subsecn:prob-prep}, the noiseless probability should be $1$, and thus the decrease from $1$ highlights the detrimental effects of the noise. While the measured Bhattacharyya distance after the Fibonacci projection remains close to $0$, the estimated energy $\langle H_J \rangle$ exhibits a linearly growing error rate as compared to the ideal value $-(N - 3)$, reaching around $10\%$ at $N=14$. Also note that the noise can create spurious samples whose $\langle H_J \rangle$ is less than the theoretical minimum of the $H_J$ in Eq.~\eqref{eq:compressed_h0}

Next, we try various error mitigation techniques to improve the estimate accuracy.
To reduce the measurement error, the iterative Bayesian unfolding \cite{Nachman2020_readout_unfolding} uses the confusion matrix of 8-bit strings that reflects Aspen-M's octagonal layout. The samples obtained from the modified bitstrings are marked as blue dots and feature interesting changes. They have a higher success probability---i.e., they remain more in the Fibonacci subspace---and 
 feature less error in estimating $\langle H_J \rangle$ after the Fibonacci projection. However, their Bhattacharyya distance also amplifies significantly with increased sample variance. Application of randomized compilation \cite{Wallman2016_randomized_compiling} and readout symmetrization \cite{Smith2021_readout_symm} can help to reduce the Bhattacharyya distance while maintaining other improvements, as seen from the orange entries in \cref{fig:sk-state-run}.

Figure \ref{fig:sk-state-tomography} displays the results of the $|\mathcal{S}_{k_\text{max}}\rangle$ state tomography after applying the aforementioned error mitigation methods and the Fibonacci post-projection. It illuminates the sparsity of the $|\mathcal{S}_{k_\text{max}}\rangle$ state that populates only $N/2$ bitstrings out of $2^{k_\text{max}}$ different choices.

\subsection{Alternative State Preparation Protocols}

Here we consider two possible alternative strategies to prepare the scarred eigenstates $\ket{\mathcal S_k}$. First, it is natural to ask whether the recursive method of Ref.~\cite{Bartschi19} to prepare Dicke states $\ket{d^m_k}$ in depth $O(mk)$ can be adapted to exactly prepare the $\ket{\mathcal S_k}$ states, or equivalently the projected Dicke states $\ket{\mathcal D^m_k}$ [see Eq.~\eqref{eq:Ddef}], in polynomial depth. The  construction of a state preparation circuit in Ref.~\cite{Bartschi19} hinges on a recursion relation between Dicke states with different magnetizations. The projected Dicke states also obey a recursion relation:
\begin{equation}
\begin{split}
    \ket{\mathcal D^{m}_k} &= \sqrt{\frac{k}{m-k+1}}\ket{\mathcal D^{m-2}_{k-1}}\otimes \ket{01}\\
    &\indent + \sqrt{\frac{m-2k+1}{m-k+1}}\ket{\mathcal D^{m-1}_k}\otimes \ket{0}.
\end{split}
\end{equation}
However, unlike the recursion relation for Dicke states, this one relates $\ket{\mathcal D^m_k}$ to both $\ket{\mathcal D^{m-1}_k}$ and $\ket{\mathcal D^{m-2}_{k-1}}$, which are defined on qubit registers of different sizes. This complicates the possibility of a straightforward modification of the protocol of Ref.~\cite{Bartschi19}, and we leave this question for future work. Furthermore, we reiterate that the $O(mk)$ depth scaling for the Dicke states is only marginally smaller in an asymptotic sense than the $O(mk\log^2k)$ scaling obtained within our MPS-to-circuit approach.

Second, it is interesting to consider nonunitary methods to prepare the scarred eigenstates $\ket{\mathcal S_k}$. In particular, these eigenstates can be prepared from the state $\ket{\xi}$ by projectively measuring the magnetization operator $M_z = \sum^N_{i=1}Z_i$. For any $k=0,\dots,N/2-1$, $\ket{\mathcal S_k}$ is an eigenstate of $M_z$ with eigenvalue $N-2k$. Since each of these eigenvalues is unique, we see from Eq.~\eqref{eq:xi_Sk} that the projection of $\ket{\xi}$ into an $M_z$ eigenspace must yield one of the $\ket{\mathcal S_k}$. The probability of a given measurement outcome $k$ is peaked around a value that depends on $|\xi|^2$; this most probable outcome can be tuned from $k=0$ as $|\xi|^2\to 0$ to $k=N/2-1$ as $|\xi|^2\to\infty$. Thus, given the ability to measure $M_z$ without fully collapsing the system into a $z$-basis eigenstate, one can in principle prepare the state $\ket{\mathcal S_k}$ in constant depth. One way to achieve such a measurement is by a dispersive coupling $\chi_d(\sum_i Z_i)a^\dagger a$ between the qubits and a single cavity mode, where the measurement outcome would be recorded as a shift of the cavity frequency by an integer multiple of $\chi_d$. A projective measurement of $M_z$ can also be achieved using an ancilla register containing $\ceil{\log_2N}$ qubits, one for each digit of the binary representation of the magnetization eigenvalue, using methods proposed in Refs.~\cite{Botelho22} and \cite{DallaTorre22}. This state preparation method also suffers from postselection overhead beyond what is needed to prepare $\ket{\xi}$: the probability of the most probable measurement outcome decays as a power law in $N$ for large $N$. However it may have the advantage of comparatively modest circuit depth relative to the unitary state preparation protocols considered here.

Finally, we mention an alternative to the above strategy based on imaginary time evolution. Since projecting $\ket{\xi}$ into an eigenspace of $M_z$ with eigenvalue $m_k = N-2k$ yields $\ket{\mathcal S_k}$, one can imagine implementing the projection by applying $e^{-\tau(M_z-m_k)^2}$ to $\ket{\xi}$ in the limit $\tau\to\infty$. This imaginary time evolution under the Hamiltonian $H_k=(M_z-m_k)^2$ can be implemented using a number of quantum algorithms~\cite{VQITE,qite_chan20,smqite,AVQITE}. This approach could be well adapted to, e.g., trapped ion devices, which can implement the nonlocal interactions contained in $H_k$ using their native all-to-all connectivity.

\section{Conclusion and Outlook}
\label{sec: conclusion}

In this work we have explored various approaches to preparing quantum many-body scar states and their superpositions on digital quantum computers with a focus on the spin-1/2 chain model of Ref.~\cite{Iadecola20}, where an emergent Fibonacci constraint (two ``1"s may not appear next to each other) ensures that superpositions of scarred eigenstates must be entangled. Linear-depth circuits to prepare a one-parameter family of area-law-entangled superpositions of scarred eigenstates, as well as a nonunitary preparation scheme that uses measurement and postselection to prepare the same superposition in constant depth are provided. We also derived an MPS representation of the tower of scarred eigenstates and used this to generate a quasipolynomial-depth circuit to prepare individual scar states; additionally, we proposed a variational scheme based on a polynomial-depth ansatz that captures the scar states with fidelity at least $99\%$ in all numerically accessible cases. Proof-of-concept demonstrations of scar-state preparation were executed on quantum hardware.

While the state preparation protocols explored in this work were formulated with a specific model in mind, they can readily be adapted to other towers of scar states. For instance, entangled superposition states analogous to $\ket{\xi}$ are known for both the AKLT model~\cite{Mark20b} and the bond-bimagnon tower in the spin-1 XY model~\cite{Schecter19,Chattopadhyay19}. In both cases, the initial state is a finite-bond-dimension MPS which can be viewed as the projection of a simple wavefunction on $\chi$-dimensional qudits. One expects that nonunitary methods along the lines of Sec.~\ref{sec: xiprep} could be used to prepare such states (see also the method of Ref.~\cite{Smith22} to prepare the AKLT ground state, which is in a similar spirit). Likewise, both towers---and many other examples of QMBS~\cite{Moudgalya20b}---admit exact (or approximate~\cite{Zhang23}) MPS representations that can be converted to quasipolynomial-depth quantum circuits~\cite{Huggins_2019,PhysRevA.101.032310}. It is also worth noting that simpler state preparation protocols can be used in some cases---for example, the bimagnon tower in the spin-1 XY model~\cite{Schecter19} can be superposed into a product state that is trivial to prepare, and the scarred eigenstates themselves are effective spin-$1/2$ Dicke states that can be generated using a slight variation on the approach of Ref.~\cite{Bartschi19}.

Our results can feed into future studies of the stability of scarred eigenstates and their dynamics on quantum computers. There are two experiments one would like to perform---one in which the state $\ket{\xi}$ is evolved under a perturbation of the Hamiltonian in Eq.~\eqref{eq:H0}, and another in which the state $\ket{\mathcal S_k}$ is evolved. In the former case, the goal is to extract the lifetime of the oscillatory scarred dynamics from the time series of a local operator's expectation value~\footnote{A version of this calculation in a case where the initial state is a product state was performed in Ref.~\cite{Chen22}.}, while in the latter it is to extract the lifetime of the scarred eigenstate from similar data. One relevant class of perturbations to consider is an external magnetic field in the $x$-direction, $\epsilon\sum_i X_i$, which breaks the conservation of the Ising domain wall number $n_{\rm DW}$ by disrupting the balance of $X$ and $ZXZ$ in the first line of Eq.~\eqref{eq:H0}~\cite{Iadecola20}. The evolution operator over a small time step $\delta t$ can be written in a Trotter product form as (setting $J=0$ for simplicity)
\begin{widetext}
\begin{align}
U(\delta t)\approx e^{-i\Delta\delta t\sum_i Z_i}\, e^{i\lambda\delta t\sum_{i\text{ odd}}Z_{i-1}X_iZ_{i+1}}\, e^{-i(\lambda+\epsilon)\delta t\sum_i X_i}\, e^{i\lambda\delta t\sum_{i\text{ even}}Z_{i-1}X_iZ_{i+1}}.
\end{align}
\end{widetext}
The nontrivial $ZXZ$ rotations above can be compiled using standard methods into four $\rm{CNOT}$ gates, two $\rm H$ gates, and an $\mathrm{R}_Z$ gate. An informative figure of merit to track while evolving the state $\ket{\xi}$ is the expectation value of, e.g., $Z_{i-1}X_iZ_{i+1}$, which exhibits coherent oscillations under evolution by $H_0$ that should decay in the presence of a nonzero $\epsilon$. For $\ket{\mathcal S_k}$, it is informative to track the expectation value of $M_z$, which is not conserved under $H_0$ for generic initial states but is conserved for the initial state $\ket{\mathcal S_k}$; thus tracking $\braket{M_z(t)}$ provides information about the fidelity decay of $\ket{\mathcal S_k}$. These calculations are in principle straightforward to carry out on present-day quantum computers, but the QPU results in this work demonstrate that a significant fraction of the QPU's coherence budget will be expended on state preparation. Thus, any near-term demonstration of this calculation must employ substantial problem-tailored optimizations and error mitigation strategies, including symmetry protection~\cite{Halimeh2023robustquantummany,Tran:2020azk}, which should be explored in future work.

\begin{acknowledgments}
This material is based upon work supported by the U.S. Department of Energy, Office of Science, National Quantum Information Science Research Centers, Superconducting Quantum Materials and Systems Center (SQMS) under the contract No.~DE-AC02-07CH11359 (E.G., A.C.Y.L., T.I., P.P.O., J.K., D.M.K.). M.S.A. acknowledges support from USRA NASA Academic Mission Services under contract No. NNA16BD14C with NASA, with this work funded under the NASA-DOE interagency agreement SAA2-403602 governing NASA’s work as part of the SQMS center, as was A.K.’s work as part of the Feynman Quantum Academy internship program at USRA.
Ames National Laboratory is operated for the U.S. Department of Energy by Iowa State University under Contract No.~DE-AC02-07CH11358 (P.P.O., T.I.).  Variational calculations and manuscript writing by A.R.~were supported by NSF Award No.~DMR-1945395. T.I.~and A.R.~acknowledge the hospitality of the Aspen Center for Physics, which is supported by NSF Grant No.~PHY-1607611. 
This research used resources of the Oak Ridge Leadership Computing Facility, which is a DOE Office of Science User Facility supported under Contract DE-AC05-00OR22725.
We acknowledge helpful discussions with Andrew Arrasmith, Emanuele Dalla Torre, Bram Evert, Pouyan Ghaemi, Lesik Motrunich, Sanjay Moudgalya, Zlatko Papi\'c, and Matt Reagor.
\end{acknowledgments}

\appendix

\setcounter{equation}{0}
\renewcommand\theequation{A\arabic{equation}}

\section{Adiabatic Preparation of $\ket{\xi}$}
\label{sec: Adiabatic}
\begin{figure}
    \centering
    \includegraphics[width=0.485\textwidth]{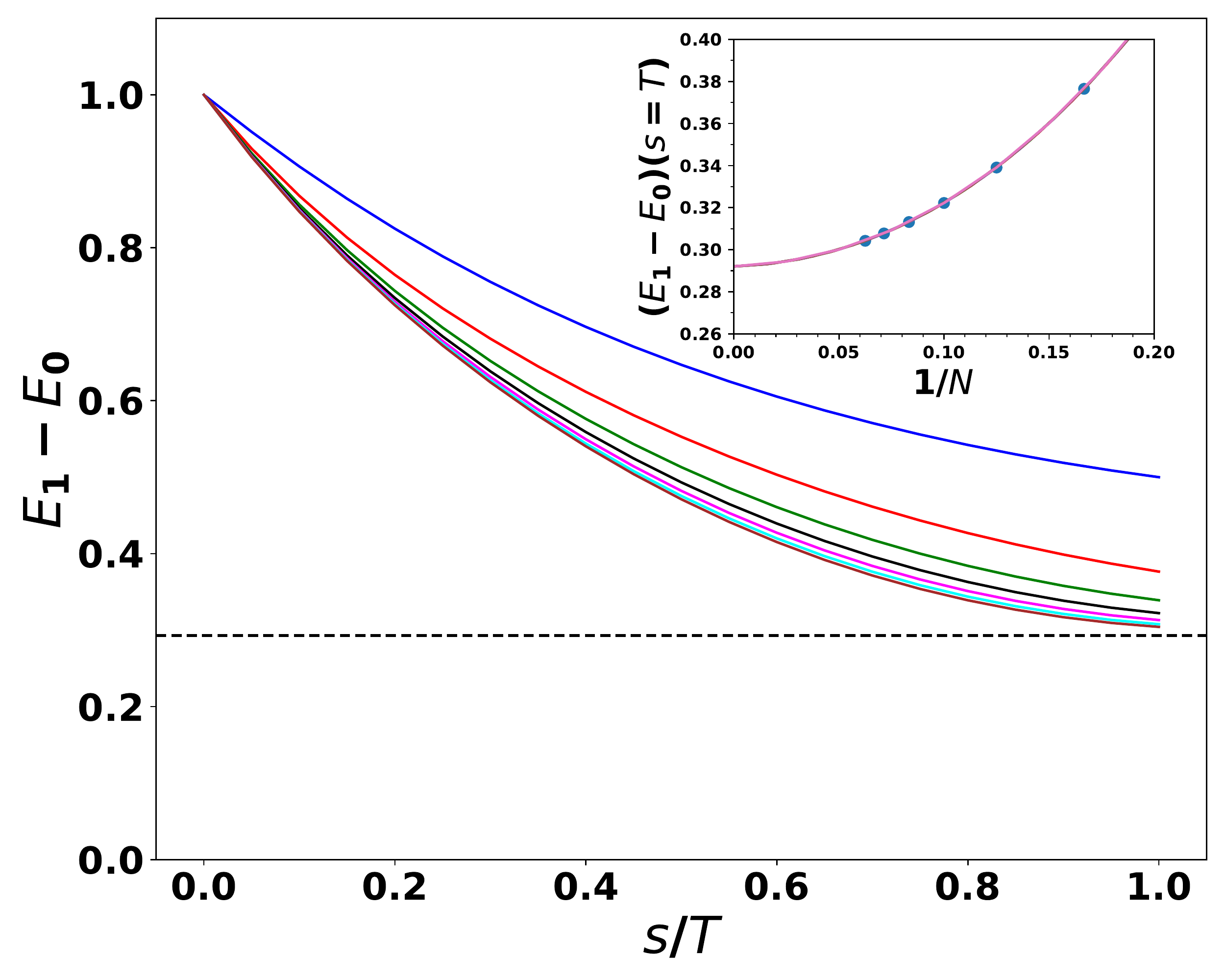}
   \caption{Energy difference between instantaneous ground and first excited states, $E_1 - E_0$, versus interpolation parameter $s/T$. The total number of spins are $N=4$ (blue), $N=6$ (red), $N=8$ (green), $N=10$ (black), $N=12$ (magenta), $N = 14$ (cyan), and $N = 16$ (brown). The minimal energy gap occurs at $s = T$ and approaches the exact value $1-1/\sqrt{2} = 0.292\dots$ (dashed horizontal line) for large $N$. This is also shown in the inset, which contains the final energy gap between the first excited and the ground state at $s = T$ versus $1/N$. The red line shows a fit to a quadratic function with $y$-intercept $0.2920$, which is consistent with the analytical value of the gap.
}
    \label{fig:energyvss}
\end{figure}
In this Appendix, we investigate the possibility of preparing the state $\ket{\xi=1}$ [Eq.~\eqref{eq: xi}] by performing a linear interpolation between the simple paramagnetic Hamiltonian 
\begin{align}
    H_X = \sum^{N-1}_{i=2}\frac{I-(-1)^i X_i}{2},
\end{align}
and the target Hamiltonian $H_P$, 
\begin{align}
\label{eq:HP}
    H_{P}=\sum^{N-1}_{i=2}\lf(P^{\prime}_{i-1}P^{\prime}_{i}+P^{\prime}_{i}P^{\prime}_{i+1}+P_{i-1}\frac{I-(-1)^i X_i}{2}P_{i+1}\rt).
\end{align}
Note that $H_P$ is simply $H_{\xi=1}/2$ [see Eq.~\eqref{eq:H_xi}], plus terms involving the projectors $P'_i=\ket 1_i\bra 1_i$ that ensure the ground state is in the Fibonacci Hilbert space.
Both Hamiltonians are positive semidefinite (i.e., their ground states have zero energy) and conserve the $Z$-basis projection of the first and last qubit, which can be fixed to be in the $0$ state.
The ground state of $H_X$ is then the product state $\ket{\psi(0)}=\ket 0 \otimes \ket{+-+-\dots+-}\otimes \ket 0$, where $\ket{\pm}$ satisfy $(1\mp X_i)\ket{\pm}=0$, while the ground state of $H_P$ is $\ket{\xi=1}$.
To adiabatically prepare the desired state, we evolve $\ket{\psi(0)}$ under the time-dependent interpolating Hamiltonian
\begin{align}
\label{eq:H(s)}
    H(s) = \lf(1-\frac{s}{T}\rt)H_X + \frac{s}{T}H_P
\end{align}
such that $H(0)=H_X$ and $H(T)=H_P$.
Provided $H(s)$ has an energy gap for all $s$, then if $T$ is sufficiently large, the time-evolved state $\ket{\psi(T)}\approx \ket{\xi=1}$ to a good approximation.
The adiabatic time evolution was carried out with exact diagonalization of the Hamiltonian~\eqref{eq:H(s)}. 
After $|\psi(0)\rangle$ is prepared, the operator $U(s, s + \delta s)$ is applied $n_s= T/(\delta s)$ times with $s = n \delta s$, $0 \leq n \leq n_s -1$. Here, $n_s$ is the number of iterations,
$T$ is the total evolution time and $\delta s$ is the time separation. The time evolution operator $U(s, s + \delta s) = e^{-i (\delta s) H(s)}$ and the overlap $|\braket{\psi(s)|\xi=1}|$ are calculated at each iteration.

\begin{figure}
    \centering
\includegraphics[width=0.485\textwidth]{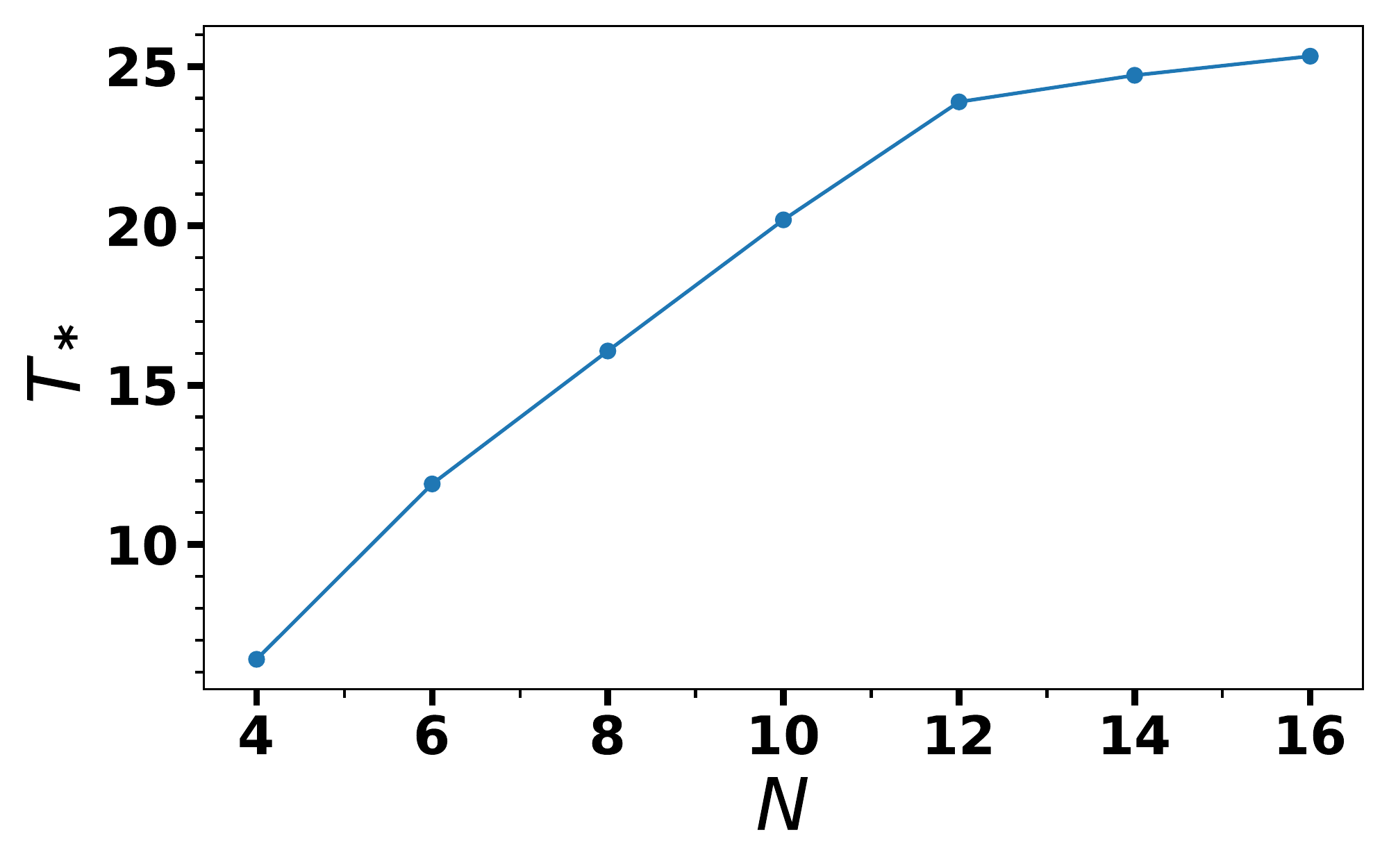}
   \caption{Total adiabatic evolution time $T_*$ needed to prepare state $\ket{\xi=1}$ with 99$\%$ fidelity as a function of system size $N$. 
   Results are obtained using exact diagonalization with a total number of steps $n_s = 10^3$
   and $\delta s = T_*/n_s$, where $ T_*$ is the minimal $T$ value to reach at least $99\%$ fidelity. 
}
    \label{fig:T*vsN}
\end{figure}

In Fig.~\ref{fig:energyvss}, we plot the difference between the instantaneous energies of the ground and first excited states $E_1 - E_0$ of $H(s)$ against the interpolation parameter, $s/T$, for several values of $N$. The minimum energy gap occurs at $s=T$ and appears to saturate to a constant as $N$ increases. In fact, an exact calculation of the energy gap of $H_{P}$ gives $1-1/\sqrt{2}= 0.292\dots$~\cite{Lesanovsky12a}, which appears to be in reasonable agreement with the limiting value of the numerically observed gap as obtained from a quadratic extrapolation in $1/N$ (see Fig.~\ref{fig:energyvss} inset). This indicates that high-fidelity adiabatic state preparation should be possible in a finite total time $T$. In Fig.~\ref{fig:T*vsN}, we plot the total time $T_*$ required to achieve state preparation with $99\%$ fidelity as a function of $N$. $T_*$ is calculated by solving for the root of the function $f(T) = |\braket{\psi(T)|\xi=1}|-0.99$ using Newton's method. We find that $T_*$ appears to approach a constant value $T_*\approx 25$. This is consistent with the behavior of the gap observed in Fig.~\ref{fig:energyvss}. 

These results lead to the conclusion that adiabatic state preparation should be possible for the state $\ket{\xi=1}$. However, it is important to compare the quantum resources required to implement adiabatic evolution on a QPU against those of the methods discussed in Sec.~\ref{sec: xiprep}. In particular, if the evolution operator is approximated using a Trotter product formula, the number of Trotter steps required to achieve Trotter error $\epsilon$ after evolution by a time $t$ scales as $t^2/\epsilon$. Inserting $t=T_*\approx 25$ and demanding a Trotter error $\epsilon=0.01$ yields an estimated $62\,500$ Trotter steps. Thus the needed circuit depth is constant, but prohibitively large for execution on present-day QPUs. This fact is a well-known shortcoming of the adiabatic algorithm~\cite{Farhi00} which partially motivated the development of alternative approaches like QAOA~\cite{Farhi14}.

\section{Derivation of Stochastic Preparation Circuit Recursion Relations}
\label{sec:stochderivation}
In this appendix, we prove the linear-depth circuit \eqref{eq:linear_preparation_circuit} preparing $\ket{\xi}$ and the associated recursion relation \eqref{eq:linear_preparation_circuit_recursive} for the preparation circuit rotation angles.

An alternative definition of $\ket{\xi}$ is given in Ref.~\cite{Iadecola20} as
\be
\ket{\xi} = \frac{1}{Z} \prod_{j=2}^{N - 1}
\lf[
1 + (-1)^{j} \xi P_{j-1} \sigma^{+}_{j} P_{j+1}
\rt]
\ket{\Omega},
\ee
where $Z$ is the normalization factor in Eq.~\eqref{eq:Z}.
Without loss of generality, we choose the convention where $j=2$ is the rightmost term in the above product, i.e., $\prod_{j=2}^{N-1} V_j \sket{\Omega} = V_{N-1} V_{N-2} \cdots V_{2} \sket{\Omega}$.
In this convention, the $(j+1)$-th qubit is always in its $\ket0$ state when applying the $j$-th term, and hence the projector $P_{j+1}$ can be omitted in the above expression. This means that
\be
\ket{\xi} = \frac{1}{Z} \prod_{j=2}^{N-1}
\lf[
1 +  (-1)^{j} \xi P_{j-1} \sigma^{+}_{j}
\rt]
\ket{\Omega}.
\ee
With the definition of $\ket{\xi;m=N-2}$ [\cref{eq:xi_m_state_definition}], we can get rid of the two qubits in the $\ket 0$ state at both ends to write
\be
\ket{\xi;m} = \frac{1}{Z}
\prod_{j=1}^{m}
\lf[
1 +  (-1)^{j+1} \xi P_{j-1} \sigma^{+}_{j}
\rt]
\ket{\Omega},
\ee
where we fix $P_{0} \equiv 1$ to be consistent with open boundary conditions.
Note that in this equation, the $j$-th qubit of $\ket{\xi;m}$ corresponds to the $(j+1)$-th qubit of $\ket{\xi}$.

To facilitate the derivation, we recall the projector $P^{\prime}_{j} =  \ket 1_j\bra 1_j = (1 - Z_j) / 2$ and introduce $x_j = (-1)^{j+1} \xi $ to rewrite the equation as
\be
\ket{\xi;m} = \frac{1}{Z} \prod_{j=1}^{m}
\lf[
P^{\prime}_{j-1} \lf(1 \rt) + P_{j-1} \lf( 1 + x_j \sigma^{+}_{j} \rt)
\rt]
\ket{\Omega},
\ee
where $P^{\prime}_{0} = 0$ for open boundary conditions. Note that to prepare the state $|\tilde\xi; m\rangle$, one can simply pick a different definition of $x_j$ to be $x_j = \xi$.

\begin{figure*}[ht!]
	\centering
	\includegraphics[width=\textwidth]{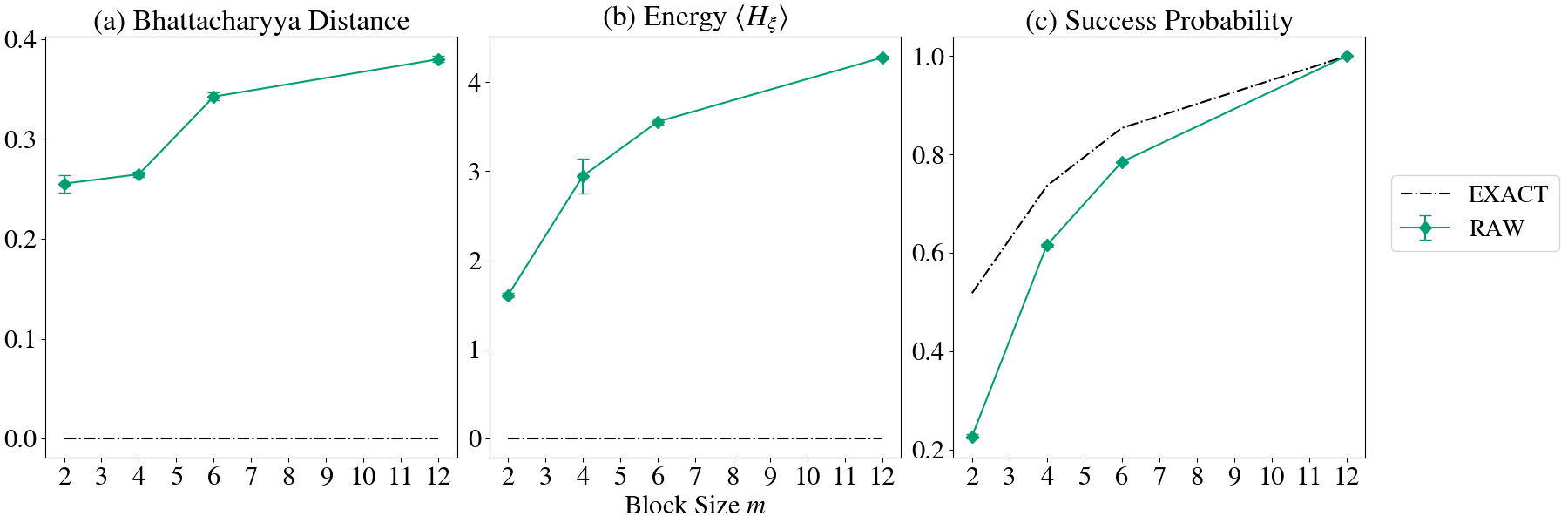}
	\caption{
		The QPU experiment results to prepare $|\xi=1\rangle$ on ``ibmq\_montreal" device with $m=2, 4, 6, 12$ and $N=14$. The $|\xi=1\rangle$ states are prepared on qubits $[26, 25, 22, 19, 16, 14, 13, 12, 10, 7, 4, 1, 2, 3]$ with ancilla qubits $[0, 6, 15, 11, 20, 24]$. We show the raw results (green diamond lines) without error mitigation to compare with the exact results (dotted line). Each data point is obtained with eight samples and $24576$ shots per sample, and the error bars denote the standard deviation over eight samples.
	}
	\label{fig:xi-state-blocksize-ibmq}
\end{figure*}

\begin{figure*}
	\centering
	\includegraphics[width=\textwidth]{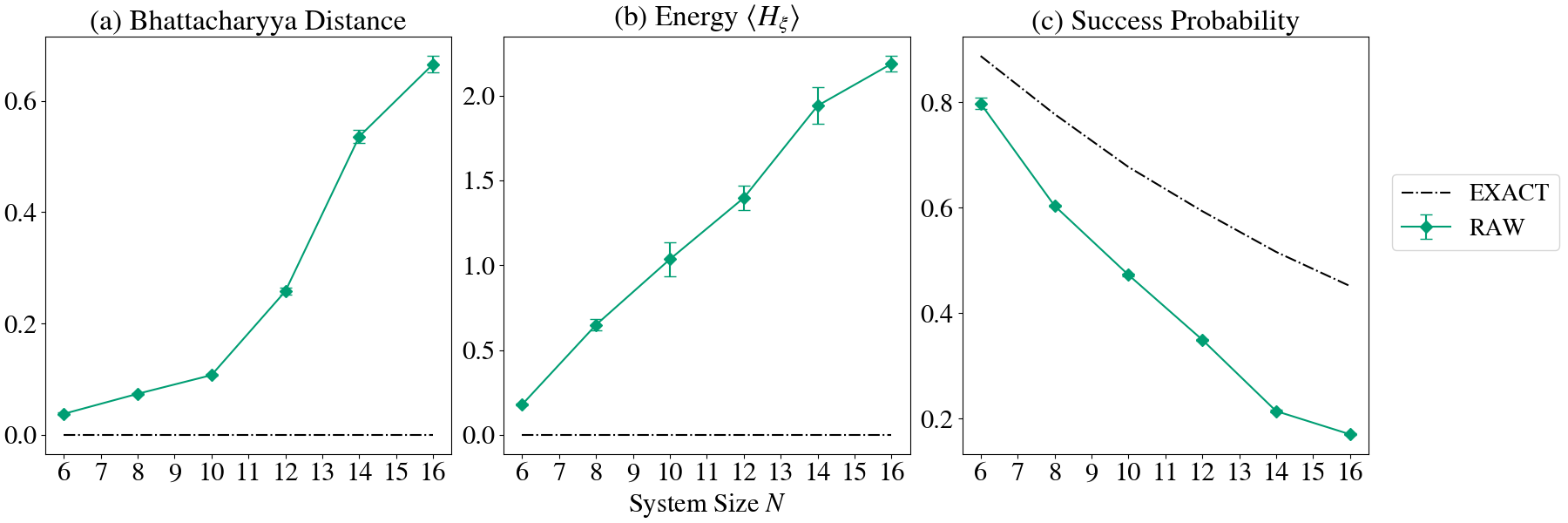}
	\caption{
		The QPU experiment results to prepare $|\xi=1\rangle$ on ``ibmq\_montreal" device with $m=2$ and $6 \le N \le 16$. The experimental setup is the same as described in \cref{fig:xi-state-blocksize-ibmq}.
	}
	\label{fig:xi-state-systemsize-ibmq}
\end{figure*}

The term in square brackets in the above equation is in the form of a controlled-rotation gate except that $( 1 + x_j  \sigma^{+}_{j} )$ is not a unitary operator.
To make it unitary, we first introduce a set of real constants $\phi_j$ whose values will be determined later in this subsection.
We can divide each term in the product by $\phi_j$ to get
\begin{equation*}
\begin{split}
\ket{\xi;m} = \frac{1}{Z'} \prod_{j=1}^{m}
[&
P^{\prime}_{j-1} (\frac{1}{\phi_j} ) + \\
&P_{j-1} \frac{1}{\phi_j}  (1 +  x_j \sigma^{+}_{j} )]\ket{\Omega},\\
\end{split}
\end{equation*}
where $Z' = Z / \prod_{j=1}^{m} \phi_j$.
We recognize that the local state $\sket{1}$ is only generated by the raising operator $\sigma^{+}_{j}$. Hence, the factor $1/\phi_j$ associated with $P^{\prime}_{j-1}$ can be moved to be associated with $\sigma^{+}_{j - 1}$ in the previous product term. We can then write
\begin{equation*}
\ket{\xi;m} = \frac{1}{Z'} \prod_{j=1}^{m}
\lf[
P^{\prime}_{j-1} + P_{j-1} \frac{1}{\phi_j} \lf( 1 + \frac{x_j }{\phi_{j+1}}  \sigma^{+}_{j} \rt)
\rt]
\ket{\Omega},
\end{equation*}
where $\phi_{m+1} = 1$.
Since $\sigma^{+}_{j}$ only acts on $\sket{0}$ in this equation, we have $\sigma^{+}_{j} \sket{0} = \mathedit{-} i \sigma^{y}_{j} \sket{0}$ where $\sigma^{y}_j = -i \ket 0_j \bra 1_j + i \ket 1_j \bra 0_j$. This allows us to write
\be
\label{eq:linear_preparation_circuit_derivation_1}
\ket{\xi;m} = \frac{1}{Z'} \prod_{j=1}^{m}
\lf[
P^{\prime}_{j-1} + P_{j-1} \frac{1}{\phi_j} \lf( 1 \mathedit{-}  i \frac{x_j}{\phi_{j+1}}  \sigma^{y}_{j} \rt)
\rt]
\ket{\Omega}.
\ee
If we pick
\be
\label{app_eq:phi_value}
\phi_j = \sqrt{1 + \lf| \frac{x_j}{\phi_{j+1}} \rt|^2} \text{\ \ and \ \ } \phi_{m + 1} = 1,
\ee
the operator associated with $P_{j-1}$ is unitary such that
\begin{equation*}
\frac{1}{\phi_j} \lf( 1 \mathedit{-}  i \frac{x_j}{\phi_{j+1}}  \sigma^{y}_{j} \rt)
= \cos\frac{\theta_j}{2} \mathedit{-}  i \sin \frac{\theta_j}{2} \sigma^{y}_{j}
= \mathrm{R}_{Y, j} (\theta_j),
\end{equation*}
where $\mathrm{R}_{Y, j}$ is the $Y$-rotation gate on the $j$-th qubit and $\theta_j = 2 \, \mathrm{arg}  ( 1 + i x_j/\phi_{j+1}  )$ is the rotation angle.
Rewriting \cref{eq:linear_preparation_circuit_derivation_1} in terms of $Y$-rotation gates, we have
\be
\ket{\xi;m} = \frac{1}{Z'} \prod_{j=1}^{m}
\lf[
P^{\prime}_{j-1} + P_{j-1}\mathrm{R}_{Y, j} (\theta_j)
\rt]
\ket{\Omega}.
\ee
Since all terms are unitary, $\frac{1}{Z'}$ is just a global phase factor not relevant for the state preparation and hence can be dropped.
Note also that $P^{\prime}_{0}=0$ and $P_{0}=1$ and hence the $j=1$ term is just a $Y$-rotation gate.
Further rewriting the above equation using the controlled-$Y$ rotation gates, we arrive at \cref{eq:linear_preparation_circuit} with $\theta_j$ given by \cref{eq:linear_preparation_circuit_recursive}.

\section{QPU results for the complete constant-depth circuit with postselection on a IBM processor}
\label{sec:ibm_result}

In this appendix, we present the QPU results of implementing the complete constant-depth circuit \eqref{eq:linear_preparation_circuit} with postselection shown in \cref{fig:quantumcircuitstitchingxi} to prepare the superposition state $\ket{\xi}$ on an IBM quantum processor without introducing the fragmentation reduction used in \cref{sec:xi-qpu}.

We compile the constant-depth and postselection circuits into native gates respecting the connectivity of the ``ibmq\_montreal" device. In particular, each controlled $Y$-rotations (Toffoli gate) is decomposed into two (10) CNOT gates and single-qubit gates. The compiled circuit consists of $2 \frac{(N - 2) (m - 1)}{m} + 10\lf(\frac{N - 2}{m} - 1\rt)$ CNOT gates and has a circuit depth of $2 (m - 1) + 10$ (counting only CNOT gates).
Similar to the experiment described in \cref{sec:xi-qpu}, we measure the Bhattacharyya distance, the expectation value of the parent Hamiltonian $H_{\xi}$ and the postselection success probability on``ibmq\_montreal" for various block size $m$ and system size $N$.
The QPU results are shown in \cref{fig:xi-state-blocksize-ibmq,fig:xi-state-systemsize-ibmq}, and each data point is measured with eight samples and 24576 shots per sample.

The performance of the state preparation protocol improves with decreasing block size $m$ indicated by the decreasing deviation between the measure values and the ideal value (0) of the Bhattacharyya distance and energy shown in \cref{fig:xi-state-blocksize-ibmq}(a) and (b). This comes with the cost of decreasing the postselection success probability shown in \cref{fig:xi-state-blocksize-ibmq}(c). The improvement is expected since the depth of the circuit decreases with decreasing $m$ even though the number of CNOT increases.
The protocol's performance deteriorates with increasing systems size $N$ as shown in \cref{fig:xi-state-systemsize-ibmq} since the circuit depth and the number of CNOT increase with $N$.
The trends in varying $m$ and $N$ discussed in this appendix are consistent with the results shown in \cref{sec:xi-qpu} using the fragmented circuits.

\bibliographystyle{quantum}
\bibliography{refs.bbl}

\end{document}